\documentclass[10pt, twocolumn, final]{IEEEtran}%
\usepackage[T1]{fontenc}
\usepackage[utf8]{inputenc}
\usepackage{mathtools}

\usepackage{amssymb,mathrsfs}% Typical maths resource packages
\usepackage{amsthm}
\usepackage{bm}
\usepackage{scalerel}
\usepackage{nicefrac}
\usepackage{microtype} 
\usepackage[shortlabels]{enumitem}
\usepackage{graphicx}
\usepackage{epstopdf}
\DeclareGraphicsExtensions{.eps,.png,.jpg,.pdf}

\usepackage{url}
\usepackage{colortbl}
\usepackage{booktabs}
\usepackage{multirow}
\usepackage{colortbl,xcolor}
\usepackage[normalem]{ulem}
\usepackage{xparse,xstring}
\usepackage{calc}
\usepackage{etoolbox}

\makeatletter
\@ifpackageloaded{natbib}{
	\relax
}{
	\usepackage{cite}
}
\makeatother

\usepackage{array}
\newcolumntype{L}[1]{>{\raggedright\let\newline\\\arraybackslash\hspace{0pt}}m{#1}}
\newcolumntype{C}[1]{>{\centering\let\newline\\\arraybackslash\hspace{0pt}}m{#1}}
\newcolumntype{R}[1]{>{\raggedleft\let\newline\\\arraybackslash\hspace{0pt}}m{#1}}

\makeatletter
\let\MYcaption\@makecaption
\makeatother
\usepackage[font=footnotesize]{subcaption}
\makeatletter
\let\@makecaption\MYcaption
\makeatother

\usepackage{glossaries}
\makeatletter
\sfcode`\.1006

\let\oldgls\gls
\let\oldglspl\glspl

\newcommand\fussy@ifnextchar[3]{%
	\let\reserved@d=#1%
	\def\reserved@a{#2}%
	\def\reserved@b{#3}%
	\futurelet\@let@token\fussy@ifnch}
\def\fussy@ifnch{%
	\ifx\@let@token\reserved@d
		\let\reserved@c\reserved@a
	\else
		\let\reserved@c\reserved@b
	\fi
	\reserved@c}

\renewcommand{\gls}[1]{%
\oldgls{#1}\fussy@ifnextchar.{\@checkperiod}{\@}}
\renewcommand{\glspl}[1]{%
\oldglspl{#1}\fussy@ifnextchar.{\@checkperiod}{\@}}

\newcommand{\@checkperiod}[1]{%
	\ifnum\sfcode`\.=\spacefactor\else#1\fi
}

\robustify{\gls}
\robustify{\glspl}
\makeatother

\newacronym{wrt}{w.r.t.}{with respect to}
\newacronym{RHS}{R.H.S.}{right-hand side}
\newacronym{LHS}{L.H.S.}{left-hand side}
\newacronym{iid}{i.i.d.}{independent and identically distributed}
\newacronym{SOTA}{SOTA}{state-of-the-art}
\usepackage{float}

\ifx\notloadhyperref\undefined
	\ifx\loadbibentry\undefined
		\usepackage[hidelinks,hypertexnames=false]{hyperref}
	\else
		\usepackage{bibentry}
		\makeatletter\let\saved@bibitem\@bibitem\makeatother
		\usepackage[hidelinks,hypertexnames=false]{hyperref}
		\makeatletter\let\@bibitem\saved@bibitem\makeatother
	\fi
\else
	\ifx\loadbibentry\undefined
		\relax
	\else
		\usepackage{bibentry}
	\fi
\fi

\usepackage[capitalize]{cleveref}
\crefname{equation}{}{}
\Crefname{equation}{}{}
\crefname{claim}{claim}{claims}
\crefname{step}{step}{steps}
\crefname{line}{line}{lines}
\crefname{condition}{condition}{conditions}
\crefname{dmath}{}{}
\crefname{dseries}{}{}
\crefname{dgroup}{}{}

\crefname{Problem}{Problem}{Problems}
\crefformat{Problem}{Problem~#2#1#3}
\crefrangeformat{Problem}{Problems~#3#1#4 to~#5#2#6}

\crefname{Theorem}{Theorem}{Theorems}
\crefname{Corollary}{Corollary}{Corollaries}
\crefname{Proposition}{Proposition}{Propositions}
\crefname{Lemma}{Lemma}{Lemmas}
\crefname{Definition}{Definition}{Definitions}
\crefname{Example}{Example}{Examples}
\crefname{Assumption}{Assumption}{Assumptions}
\crefname{Remark}{Remark}{Remarks}
\crefname{Rem}{Remark}{Remarks}
\crefname{remarks}{Remarks}{Remarks}
\crefname{Appendix}{Appendix}{Appendices}
\crefname{Supplement}{Supplement}{Supplements}
\crefname{Exercise}{Exercise}{Exercises}
\crefname{Theorem_A}{Theorem}{Theorems}
\crefname{Corollary_A}{Corollary}{Corollaries}
\crefname{Proposition_A}{Proposition}{Propositions}
\crefname{Lemma_A}{Lemma}{Lemmas}
\crefname{Definition_A}{Definition}{Definitions}

\usepackage{crossreftools}
\ifx\notloadhyperref\undefined
\else
	\relax
\fi

\usepackage{algorithm}
\usepackage{algpseudocode}

\ifx\loadbreqn\undefined
	\relax
\else
	\usepackage{breqn}
\fi

\makeatletter
\def\cleartheorem#1{%
    \expandafter\let\csname#1\endcsname\relax
    \expandafter\let\csname c@#1\endcsname\relax
}
\def\clearthms#1{ \@for\tname:=#1\do{\cleartheorem\tname} }
\makeatother

\ifx\renewtheorem\undefined
	\ifx\useTheoremCounter\undefined
		\newtheorem{Theorem}{Theorem}
		\newtheorem{Corollary}{Corollary}
		\newtheorem{Proposition}{Proposition}
		\newtheorem{Lemma}{Lemma}
	\else
		\newtheorem{Theorem}{Theorem}
		
		\newtheorem{Proposition}[Theorem]{Proposition}
	\fi

	\newtheorem{Definition}{Definition}
	\newtheorem{Example}{Example}

\fi

\theoremstyle{remark}

\theoremstyle{plain}

\newcommand{\qednew}{\nobreak \ifvmode \relax \else
		\ifdim\lastskip<1.5em \hskip-\lastskip
			\hskip1.5em plus0em minus0.5em \fi \nobreak
		\vrule height0.75em width0.5em depth0.25em\fi}



\NewDocumentCommand{\movedownsub}{e{^_}}{%
	\IfNoValueTF{#1}{%
		\IfNoValueF{#2}{^{}}% neither ^ nor _, do nothing; if no ^ but _, add ^{}
	}{%
		^{#1}% add superscript if present
	}%
	\IfNoValueF{#2}{_{#2}}% add subscript if present
}

\let\latexchi\chi
\RenewDocumentCommand{\chi}{}{\latexchi\movedownsub}

\newcommand{\Real}{\mathbb{R}}

\newcommand{\Complex}{\mathbb{C}}

\newcommand{\calE}{\mathcal{E}}

\newcommand{\bB}{\mathbf{B}}

\newcommand{\be}{\mathbf{e}}
\newcommand{\bE}{\mathbf{E}}
\newcommand{\boldf}{\mathbf{f}}
\newcommand{\bF}{\mathbf{F}}
\newcommand{\bg}{\mathbf{g}}

\newcommand{\bL}{\mathbf{L}}

\newcommand{\bM}{\mathbf{M}}

\newcommand{\bP}{\mathbf{P}}

\newcommand{\bu}{\mathbf{u}}

\newcommand{\bw}{\mathbf{w}}

\newcommand{\bx}{\mathbf{x}}

\newcommand{\by}{\mathbf{y}}

\newcommand{\bz}{\mathbf{z}}

\newcommand{\bbR}{\mathbb{R}}

\newcommand{\bbZ}{\mathbb{Z}}

\DeclareSymbolFont{bsfletters}{OT1}{cmss}{bx}{n}
\DeclareSymbolFont{ssfletters}{OT1}{cmss}{m}{n}
\DeclareMathSymbol{\bsfGamma}{0}{bsfletters}{'000}
\DeclareMathSymbol{\ssfGamma}{0}{ssfletters}{'000}
\DeclareMathSymbol{\bsfDelta}{0}{bsfletters}{'001}
\DeclareMathSymbol{\ssfDelta}{0}{ssfletters}{'001}
\DeclareMathSymbol{\bsfTheta}{0}{bsfletters}{'002}
\DeclareMathSymbol{\ssfTheta}{0}{ssfletters}{'002}
\DeclareMathSymbol{\bsfLambda}{0}{bsfletters}{'003}
\DeclareMathSymbol{\ssfLambda}{0}{ssfletters}{'003}
\DeclareMathSymbol{\bsfXi}{0}{bsfletters}{'004}
\DeclareMathSymbol{\ssfXi}{0}{ssfletters}{'004}
\DeclareMathSymbol{\bsfPi}{0}{bsfletters}{'005}
\DeclareMathSymbol{\ssfPi}{0}{ssfletters}{'005}
\DeclareMathSymbol{\bsfSigma}{0}{bsfletters}{'006}
\DeclareMathSymbol{\ssfSigma}{0}{ssfletters}{'006}
\DeclareMathSymbol{\bsfUpsilon}{0}{bsfletters}{'007}
\DeclareMathSymbol{\ssfUpsilon}{0}{ssfletters}{'007}
\DeclareMathSymbol{\bsfPhi}{0}{bsfletters}{'010}
\DeclareMathSymbol{\ssfPhi}{0}{ssfletters}{'010}
\DeclareMathSymbol{\bsfPsi}{0}{bsfletters}{'011}
\DeclareMathSymbol{\ssfPsi}{0}{ssfletters}{'011}
\DeclareMathSymbol{\bsfOmega}{0}{bsfletters}{'012}
\DeclareMathSymbol{\ssfOmega}{0}{ssfletters}{'012}

\newcommand{\bomega}{\bm{\omega}}

\makeatletter
\newcommand*\rel@kern[1]{\kern#1\dimexpr\macc@kerna}
\newcommand*\widebar[1]{%
  \begingroup
  \def\mathaccent##1##2{%
    \rel@kern{0.8}%
    \overline{\rel@kern{-0.8}\macc@nucleus\rel@kern{0.2}}%
    \rel@kern{-0.2}%
  }%
  \macc@depth\@ne
  \let\math@bgroup\@empty \let\math@egroup\macc@set@skewchar
  \mathsurround\z@ \frozen@everymath{\mathgroup\macc@group\relax}%
  \macc@set@skewchar\relax
  \let\mathaccentV\macc@nested@a
  \macc@nested@a\relax111{#1}%
  \endgroup
}
\makeatother

\DeclareMathOperator{\ST}{s.t.\ }

\DeclareMathOperator{\var}{var}

\DeclareMathOperator{\cov}{cov}

\DeclareMathOperator{\rank}{rank}

\DeclareMathOperator{\ima}{im}

\newcommand{\ifbcdot}[1]{\ifblank{#1}{\cdot}{#1}}

\DeclarePairedDelimiterX\abs[1]{\lvert}{\rvert}{\ifbcdot{#1}}
\DeclarePairedDelimiterX\parens[1]{(}{)}{\ifbcdot{#1}}
\DeclarePairedDelimiterX\brk[1]{[}{]}{\ifbcdot{#1}}
\DeclarePairedDelimiterX\braces[1]{\{}{\}}{\ifbcdot{#1}}
\DeclarePairedDelimiterX\angles[1]{\langle}{\rangle}{\ifblank{#1}{\cdot,\cdot}{#1}}
\DeclarePairedDelimiterX\ip[2]{\langle}{\rangle}{\ifbcdot{#1},\ifbcdot{#2}}
\DeclarePairedDelimiterX\norm[1]{\lVert}{\rVert}{\ifbcdot{#1}}
\DeclarePairedDelimiterX\ceil[1]{\lceil}{\rceil}{\ifbcdot{#1}}
\DeclarePairedDelimiterX\floor[1]{\lfloor}{\rfloor}{\ifbcdot{#1}}

\DeclareFontFamily{U}{matha}{\hyphenchar\font45}
\DeclareFontShape{U}{matha}{m}{n}{
      <5> <6> <7> <8> <9> <10> gen * matha
      <10.95> matha10 <12> <14.4> <17.28> <20.74> <24.88> matha12
      }{}
\DeclareSymbolFont{matha}{U}{matha}{m}{n}
\DeclareFontSubstitution{U}{matha}{m}{n}

\DeclareFontFamily{U}{mathx}{\hyphenchar\font45}
\DeclareFontShape{U}{mathx}{m}{n}{
      <5> <6> <7> <8> <9> <10>
      <10.95> <12> <14.4> <17.28> <20.74> <24.88>
      mathx10
      }{}
\DeclareSymbolFont{mathx}{U}{mathx}{m}{n}
\DeclareFontSubstitution{U}{mathx}{m}{n}

\DeclareMathDelimiter{\vvvert}{0}{matha}{"7E}{mathx}{"17}
\DeclarePairedDelimiterX\vertiii[1]{\vvvert}{\vvvert}{\ifbcdot{#1}}

\DeclarePairedDelimiterXPP\trace[1]{\operatorname{Tr}}{(}{)}{}{\ifbcdot{#1}} % column vector
\DeclarePairedDelimiterXPP\col[1]{\operatorname{col}}{\{}{\}}{}{\ifbcdot{#1}} % column vector
\DeclarePairedDelimiterXPP\row[1]{\operatorname{row}}{\{}{\}}{}{\ifbcdot{#1}} % row vector
\DeclarePairedDelimiterXPP\erf[1]{\operatorname{erf}}{(}{)}{}{\ifbcdot{#1}}
\DeclarePairedDelimiterXPP\erfc[1]{\operatorname{erfc}}{(}{)}{}{\ifbcdot{#1}}
\DeclarePairedDelimiterXPP\KLD[2]{D}{(}{)}{}{\ifbcdot{#1}\, \delimsize\|\, \ifbcdot{#2}} % KL divergence
\DeclarePairedDelimiterXPP\op[2]{\operatorname{#1}}{(}{)}{}{#2} % general operator

\newcommand{\T}{^{\mkern-1.5mu\mathop\intercal}}% transpose notation
\newcommand{\ud}{\,\mathrm{d}} % for integrals like \int f(x) \ud x
\DeclarePairedDelimiterXPP\indicate[1]{{\bf 1}}{\{}{\}}{}{\ifbcdot{#1}}

\NewDocumentCommand\ofrac{s m}{%
	\IfBooleanTF#1%
	{\dfrac{1}{#2}}%
	{\frac{1}{#2}}%
}
\NewDocumentCommand\ddfrac{s m m}{%
	\IfBooleanTF#1%
	{\dfrac{\mathrm{d} {#2}}{\mathrm{d} {#3}}}%
	{\frac{\mathrm{d} {#2}}{\mathrm{d} {#3}}}%
}
\NewDocumentCommand\ppfrac{s m m}{%
	\IfBooleanTF#1%
	{\dfrac{\partial {#2}}{\partial {#3}}}%
	{\frac{\partial {#2}}{\partial {#3}}}%
}

\providecommand\given{}
\DeclarePairedDelimiterX\Set[2]\{\}{%
\renewcommand\given{\SetSymbol[\delimsize]{#1}}
#2
}
\DeclarePairedDelimiterX\Setc[1]\{\}{%
\renewcommand\given{\SetSymbol{:}}
#1
}

\NewDocumentCommand\set{s o m}{%
	\IfBooleanTF#1%
	{\IfValueTF{#2}{\Set*{#2}{#3}}{\Setc*{#3}}}%
	{\IfValueTF{#2}{\Set{#2}{#3}}{\Setc{#3}}}%
}

\NewDocumentCommand{\evalat}{ s O{\big} m e{_^} }{%
\IfBooleanTF{#1}%
{\left. #3 \right|}{#3#2|}%
\IfValueT{#4}{_{#4}}%
\IfValueT{#5}{^{#5}}%
}

\providecommand\given{}
\DeclarePairedDelimiterXPP\cprob[1]{}(){}{
\renewcommand\given{\nonscript\,\delimsize\vert\allowbreak\nonscript\,\mathopen{}}%
#1%
}
\DeclarePairedDelimiterXPP\cexp[1]{}[]{}{
\renewcommand\given{\nonscript\,\delimsize\vert\allowbreak\nonscript\,\mathopen{}}%
#1%
}

\DeclareDocumentCommand \P { s e{_^} d() g } {%
	\mathbb{P}%
	\IfBooleanTF{#1}%
		{
			\IfValueT{#2}{_{#2}}%
			\IfValueT{#3}{^{#3}}%
			\IfValueTF{#5}{\cprob{#4 \given #5}}{\IfValueT{#4}{\cprob{#4}}}%
		}%
		{
			\IfValueT{#2}{_{#2}}%
			\IfValueT{#3}{^{#3}}%
			\IfValueTF{#5}{\cprob*{#4 \given #5}}{\IfValueT{#4}{\cprob*{#4}}}%
		}%
}

\DeclareDocumentCommand \E { s e{_^} o g } {%
	\mathbb{E}%
	\IfBooleanTF{#1}%
		{
			\IfValueT{#2}{_{#2}}%
			\IfValueT{#3}{^{#3}}%
			\IfValueTF{#5}{\cexp{#4 \given #5}}{\IfValueT{#4}{\cexp{#4}}}%
		}%
		{
			\IfValueT{#2}{_{#2}}%
			\IfValueT{#3}{^{#3}}%	
			\IfValueTF{#5}{\cexp*{#4 \given #5}}{\IfValueT{#4}{\cexp*{#4}}}%		
		}%
}

\DeclareDocumentCommand \Var { s e{_^} d() g } {%
	\var%
	\IfBooleanTF{#1}%
		{
			\IfValueT{#2}{_{#2}}%
			\IfValueT{#3}{^{#3}}%
			\IfValueTF{#5}{\cprob{#4 \given #5}}{\IfValueT{#4}{\cprob{#4}}}%
		}%
		{
			\IfValueT{#2}{_{#2}}%
			\IfValueT{#3}{^{#3}}%	
			\IfValueTF{#5}{\cprob*{#4 \given #5}}{\IfValueT{#4}{\cprob*{#4}}}%		
		}%
}

\DeclareDocumentCommand \Cov { s e{_^} d() g } {%
	\cov%
	\IfBooleanTF{#1}%
		{
			\IfValueT{#2}{_{#2}}%
			\IfValueT{#3}{^{#3}}%
			\IfValueTF{#5}{\cprob{#4 \given #5}}{\IfValueT{#4}{\cprob{#4}}}%
		}%
		{
			\IfValueT{#2}{_{#2}}%
			\IfValueT{#3}{^{#3}}%	
			\IfValueTF{#5}{\cprob*{#4 \given #5}}{\IfValueT{#4}{\cprob*{#4}}}%		
		}%
}

\ExplSyntaxOn
\NewDocumentCommand \dist {m o o} {%
\mathrm{#1}\left(%
	\IfValueT{#3}{%
		\tl_if_blank:nTF{ #3 }{\cdot\, \middle|\, }{#3\, \middle|\, }%
	}
	\IfValueT{#2}{#2}%
\right)%
}
\ExplSyntaxOff

\NewDocumentCommand {\cbrace} {t+ D[]{black} D(){\widthof{#5}} m m } {%
	\begingroup%
		\color{#2}
		\IfBooleanTF{#1}{%
			\overbrace{#4}^%
		}{
			\underbrace{#4}_%
		}%
		{\parbox[c]{#3}{\centering\footnotesize{#5}}}%
	\endgroup% 
}

\let\oldforall\forall
\renewcommand{\forall}{\oldforall \, }

\let\oldexist\exists
\renewcommand{\exists}{\oldexist \, }

\makeatletter

\newcommand{\rankcolor}[2]{%
	\expandafter\renewcommand\csname #1\endcsname[1]{%
		\ifblank{##1}{%
			{\color{#2} \textbf{#2}}%
		}{%
			\ifmmode
				\textcolor{#2}{\bm{##1}}%
			\else%
				{\color{#2} \textbf{##1}}%
			\fi	
		}%
	}
}

\rankcolor{first}{red}
\rankcolor{second}{blue}
\rankcolor{third}{cyan}
\makeatother

\newcommand{\figref}[1]{Fig.~\ref{#1}}
\graphicspath{{./Figures/}{./figures/}}
\DeclareDocumentCommand{\includeCroppedPdf}{ o O{./Figures/} m }{
	\IfFileExists{#2#3-crop.pdf}{}{%
		\immediate\write18{pdfcrop #2#3.pdf #2#3-crop.pdf}}%
	\includegraphics[#1]{#2#3-crop.pdf}
}

\makeatletter
\newcommand*{\addFileDependency}[1]{% argument=file name and extension
  \typeout{(#1)}
  \@addtofilelist{#1}
  \IfFileExists{#1}{}{\typeout{No file #1.}}
}
\makeatother

\definecolor{gray90}{gray}{0.9}
\def\colorlist{red,blue,brown,cyan,darkgray,gray,lightgray,green,lime,magenta,olive,orange,pink,purple,teal,violet,white,yellow}

\makeatletter
\def\startcomment{[}
\ifx\nohighlights\undefined
	\newcommand{\createcolor}[1]{%
			\expandafter\newcommand\csname #1\endcsname[1]{{\color{#1} ##1}}%
	}
	\newcommand{\msout}[1]{\text{\color{green} \sout{\ensuremath{#1}}}}
	\newcommand{\del}[1]{{\color{green}\ifmmode \msout{#1}\else\sout{#1}\fi}}
\else
	\newcommand{\createcolor}[1]{%
			\expandafter\newcommand\csname #1\endcsname[1]{%
				\noexpandarg%
				\StrChar{##1}{1}[\firstletter]%
				\if\firstletter\startcomment%
					\relax
				\else%
					##1
				\fi
			}%
	}
	\newcommand{\msout}[1]{}
	\newcommand{\del}[1]{}
\fi

\def\@tempa#1,{%
    \ifx\relax#1\relax\else
        \createcolor{#1}%
        \expandafter\@tempa
    \fi
}
\expandafter\@tempa\colorlist,\relax,
\makeatother

\newcommand{\hhide}[1]{}
\ifx\diagnoselabel\undefined
	\relax
\else
	\makeatletter
	\def\@testdef #1#2#3{%
		\def\reserved@a{#3}\expandafter \ifx \csname #1@#2\endcsname
			\reserved@a  \else
			\typeout{^^Jlabel #2 changed:^^J%
				\meaning\reserved@a^^J%
				\expandafter\meaning\csname #1@#2\endcsname^^J}%
			\@tempswatrue \fi}
	\makeatother
\fi

\newif\ifarxiv
\arxivtrue

\crefname{question}{question}{questions}

\usepackage{tikz}
\usepackage{pgfplots}
\pgfplotsset{compat=1.5}%%%%%%%%%%%%%%%%%%%

\providecommand{\U}[1]{\protect\rule{.1in}{.1in}}

\theoremstyle{definition}

\newacronym{GGSP}{GGSP}{generalized graph signal processing}
\newacronym{GSP}{GSP}{graph signal processing}
\newacronym{TSP}{TSP}{topological signal processing}

\title{Generalized Signals on Simplicial Complexes}
\author{Feng~Ji, Xingchao~Jian, Wee Peng Tay~\IEEEmembership{Senior Member,~IEEE}, Maosheng~Yang
\thanks{F.\ Ji, X.\ Jian, W.\ P.\ Tay are with the School of Electrical and Electronic Engineering, Nanyang Technological University, 639798, Singapore. M. Yang is with TU Delft, Netherlands. (e-mail: jifeng@ntu.edu.sg, xingchao001@e.ntu.edu.sg, wptay@ntu.edu.sg)}}
\begin{document}

\maketitle

\begin{abstract}
   Topological signal processing (TSP) over simplicial complexes typically assumes observations associated with the simplicial complexes are real scalars. In this paper, we develop TSP theories for the case where observations belong to general abelian groups, including function spaces that are commonly used to represent time-varying signals. Our approach generalizes the Hodge decomposition and allows for signal processing tasks to be performed on these more complex observations. We propose a unified and flexible framework for TSP that expands its applicability to a wider range of signal processing applications. Numerical results demonstrate the effectiveness of this approach and provide a foundation for future research in this area.
\end{abstract}


\section{Introduction} \label{sec:intro}
In recent years, there has been growing interest in analyzing signals supported on different domains. For signals supported on vertices of graphs, \gls{GSP} has emerged as a powerful tool \cite{OrtFroKov:J18, LeuMarMou:J23, JiaFenTay:J23}. For signals supported on edges of simplicial complexes such as in telecommunication traffic flows, \gls{TSP} is developed \cite{BarSar:J20, SarBarTes:C21,BarSar:J20M, SarBar:C22}. The basic idea behind \gls{TSP} is based on the Hodge decomposition of signals, which decomposes a signal into three orthogonal components: the irrotational, solenoidal, and harmonic components. Like \gls{GSP}, this decomposition is also based on the eigenspaces of the Laplacian operator and enables the consideration of concepts such as frequency, Fourier transform, wavelets, and convolution \cite{RodFraSch:C22, YanIsuSch:J22}.

These works have largely focused on the case when the signal on each edge is a scalar. To deal with discrete time-varying signals, \cite{RodGraFra:C23} has proposed an approach that embeds the signals in a product space, similar to the time-vertex framework in GSP \cite{PerLouGraVan:C17, GraLouPer:J18, LouPer:J19}. By leveraging smoothness over both spatial and temporal domains, this method is able to effectively reconstruct time-varying flows. Using a similar strategy, \cite{MonKriBef:J23} proposed an online imputation method for edge flows.

In practice, the signal space can be even more complex, e.g., a general example is given by $L^2(\Omega)$ where $\Omega$ is a continuous time interval, which allows the modeling of continuous-time signals not captured by the frameworks of \cite{RodGraFra:C23, MonKriBef:J23}. For a simplicial complex modeling a network like a sensor network (cf.\ \cite[Section V.]{Ji19}), observations or samples collected at each simplex may not be time synchronous. If our signal model uses only real-valued signals, then we need to consider multiple discrete timestamps and process observed signals on the entire network either separately for each timestamp \cite{BarSar:J20} or jointly \cite{RodGraFra:C23}. As observations are asynchronous, at each discrete timestamp, the available information can be highly incomplete. Moreover, asynchronism may render uniform discretization difficult. On the other hand, considering a function as a signal alleviates such problems by aggregating several synchronous observations into a single piece of information that is aligned for different simplexes (cf.\ \cref{sec:hfb}). Another example to consider a general signal space is in the construction of linear codes using graphs \cite{Sip96,Bom07}, when bits are usually used for edge signals. A bit has only two elements $\{0,1\}$ and has the exotic addition that $1+1=0$, as compared with the usual addition of numbers. To describe such signals, we also need an algebraic entity different from $\bbR$. 

In this paper, we develop a generalized framework that extends the existing signal processing approach to include all cases where the observations on simplexes come from a general abelian group. This encompasses a wide range of observation spaces, including those from complex numbers $\Complex$, real numbers $\Real$, and integers $\bbZ$, polynomial ring, $L^2(\Omega)$, etc., thus providing a more general and unified signal processing framework. 

To deal with these general signals, we propose to adopt the well-established setup of algebraic topology \cite{Hat02}. For example, we may use abelian groups or modules to describe signals on each simplex, denoted by $A$. Classically, $A=\bbR$. Therefore, signals on a simplicial complex $X$ are nothing but chains (of the appropriate dimension) with coefficients in $A$. In this case, many constructions in TSP \cite{BarSar:J20} require revamping.  
For example, many linear algebraic operations, such as the Hodge decomposition and Fourier transform, may not exist for a general $A$ that is not a field such as $\bbR$. It is our goal to show that by combining algebraic topology and optimization, we may still perform most (classical) signal processing tasks like filtering and reconstruction for a general $A$. 

We summarize the main contributions as follows:
\begin{itemize}
    \item We introduce generalized signals as elements from a general abelian group or modules. We illustrate how such signals can arise in practice. 
    \item Combining algebraic topology and optimization, we propose a fundamental learning framework, that generalizes classical operations such as the Hodge decomposition.
    \item We study filters and related constructions for generalized signals. 
    \item We describe various scenarios with case studies, where considering $A\neq \bbR$ can be beneficial. We demonstrate the implementation of the framework with real examples such as heat flow reconstruction in an indoor environment and analysis of the currency exchange market. 
\end{itemize}

The rest of this paper is organized as follows. In \cref{sec:prelim}, we provide an introduction to simplicial complexes and the fundamentals of TSP. In \cref{sec:alg_theory}, we give a self-contained introduction to abstract algebraic concepts such as abelian groups, rings, and modules. We can thus study the case where signals on simplexes are elements from a general abelian group or module (over a base ring). We highlight that in this case, the harmonic component should be replaced by elements of the homology group, which may not coincide with the kernel of Laplacian in general. In \cref{sec:learning}, we propose a learning framework that leverages the generalized TSP approach to perform signal processing tasks such as decomposition and imputation. In \cref{sec:fil}, we describe filters of types not commonly encountered in GSP and TSP. In \cref{sec:exp}, we present numerical results to illustrate the implementation of the proposed framework and demonstrate its usefulness. We conclude in \cref{sec:conc}. The proofs are in the Appendix.

\emph{Notations.} We use plain lower cases (e.g., $x$) to represent scalars and scalar-valued functions. We use bold lowercase (e.g., $\bx$) to represent vectors and vector-valued functions. For points and other simplexes that are geometric objects, we still use plain lowercase to write them though they may be embedded in a vector space. The bold upper cases (e.g., $\bB$) are used to denote matrices. Simplicial complexes are usually denoted by $X, Y$, and graphs are usually denoted by $G$. A general abelian group is denoted by $A$. We use $\partial_k$ to denote the $k$-th boundary operator of a simplicial complex $X$. We may also write subscripts $_{k,A}$ or $_{k,A,X}$ (for maps) on rare occasions to highlight the group $A$ and the simplicial complex $X$. We use $\T$ to denote matrix transpose. The set of non-negative numbers is deonted by $\Real_+$.

\section{Preliminaries}\label{sec:prelim}
In this section, we give a brief self-contained overview of the theory of simplicial complexes and TSP. We refer interested readers to \cite{Spa66, Hat02, BarSar:J20} for more details.  

\subsection{Simplicial complexes} \label{sec:sco}
Simplicial complexes are high-dimensional generalizations of graphs.   

\begin{Definition} \label{defn:tss}
The \emph{standard $n$-simplex (or dimension $n$ simplex)} is defined as the set
\begin{align*}
\Delta_n = \big\{(r_0,\ldots, r_n) \in \bbR_+^{n+1} \given \sum_{i=0}^n r_i=1\big\}. 
\end{align*}
Each point $v_i\in \bbR_+^{n+1}$ with its $i$-th component $1$ is a vertex of $\Delta_n$. A topological space homeomorphic to the standard $n$-simplex is called an $n$-simplex. In $\Delta_n$, if we require $k>0$ coordinates being $0$, we get an $(n-k)$-simplex, called a \emph{face}. 

A \emph{simplicial complex} $X$ is a set of simplexes such that
\begin{itemize}
    \item Any face from a simplex of $X$ is also in $X$.
    \item The intersection of any two simplexes $\sigma_1,\sigma_2\in X$ is a face of both $\sigma_1$ and $\sigma_2$.
\end{itemize} 
A simplex of $X$ is called \emph{maximal} if it is not the face of any other simplexes. 
\end{Definition}

We primarily focus on finite simplicial complexes, which have finitely many simplexes. The dimension $\dim X$ of $X$ is the largest dimension of a simplex in $X$. An $k$-dimensional simplicial complex is called a $k$-complex. For example, a $1$-complex is a graph in the traditional sense.  

Combinatorially, if we do not want to specify an exact homeomorphism of a $n$-simplex $X$ with $\Delta_n$, we represent it by $n+1$ labels of its vertices $\{v_0,\ldots,v_n\}$. Therefore, its faces are just subsets of the labels. It is worth pointing out that according to the above definition, a simplicial complex is a set of topological spaces, each homeomorphic to a simplex and they are related to each other by face relations. It is possible to produce a concrete geometric object for each simplicial complex. The \emph{geometric realization} $|X|$ of a simplicial complex $X$ is the topological space obtained by gluing simplexes with common faces. In this paper, we usually do not differentiate a simplicial complex from its geometric realization.

\subsection{Topological signal processing} \label{sec:top}
In traditional GSP, for a graph $G=(V,E)$ of size $|V|=n$ with $V=\set{v_1,\dots,v_n}$, a graph signal $\bx = (x_1,\dots,x_n) \in \bbR^n$ assigns a value $x_i$ to the node $v_i$. \gls{TSP} is the generalization of this setting. In the following content, we introduce the fundamental elements of \gls{TSP}.

Suppose $\sigma^k = \{v_0, \ldots, v_k\}$ is a $k$-simplex. An \emph{orientation} of $\sigma^k$ is an ordering of the vertices of $\sigma^k$: $[v_{i_0}, \ldots, v_{i_k}]$ with $\{i_0,\ldots, i_k\} = \{0,\ldots, k\}$. We remark that the notation $\{v_0, \ldots, v_k\}$ disregards the ordering of its elements. However, $[v_{i_0}, \ldots, v_{i_k}]$ emphasizes the ordering of the vertices. Two orientations are the same if they differ by an \emph{even permutation} and the opposite if they differ by an \emph{odd permutation}. For example, $[v_0, v_1, v_2]$ has the same orientation as $[v_1, v_2, v_0]$, while it is opposite to $[v_1, v_0, v_2]$.  Opposite orientations of the same simplex differ by a factor of $-1$. 

Assume a finite simplicial complex $X$ has $n_k$ oriented $k$-simplexes $\{\sigma^k_1,\ldots,\sigma^k_{n_k}\}$. Let the space of \emph{$k$-chains} $C_k(X,\bbR)$ be the \emph{free vector space} with basis $\{\sigma^k_1,\ldots,\sigma^k_{n_k}\}$, i.e., each $k$-chain is of a formal sum 
\begin{align*}
\sum_{1\leq i\leq n_k}r_i\sigma^k_i,\ r_i\in \bbR\ \text{for all}\ i.
\end{align*}
An equivalent interpretation is that $\sum_{1\leq i\leq n_k}r_i\sigma^k_i$ assigns the value $r_i$ to $\sigma_i^k$. Hence, $C_k(X,\bbR)$ is also understood as the space of signals on $k$-simplexes. As in traditional GSP, $C_k(X,\bbR)$ can be identified with $\bbR^{n_k}$ via the isomorphism $\sum_{1\leq i\leq n_k}r_i\sigma^k_i \mapsto (r_1,\dots,r_{n_k})$. A key player in the theory is the \emph{boundary operator} $\partial_k: C_k(X,\bbR)\to C_{k-1}(X,\bbR)$. As $C_k(X,\bbR)$ is a vector space spanned by $\{\sigma^k_1,\ldots,\sigma^k_{n_k}\}$, it suffices to define $\partial_k$ on these basis elements. 

\begin{Definition} \label{defn:ssi}
    Suppose $\sigma_i^k$ has orientation $[v_{i_0},\ldots,v_{i_k}]$. Then \begin{align*}\partial_k[v_{i_0},\ldots,v_{i_k}] = \sum_{j=0}^{k}(-1)^j[v_{i_0},\ldots,v_{i_{j-1}},v_{i_{j+1}},\ldots, v_k].\end{align*} 
\end{Definition}

For example, if $\sigma^1 = [v_0,v_1]$ is a directed edge with edge signal $r$, $\partial_1(r\sigma^1) = r[v_1]+(-r)[v_0]$, i.e., $r$ is assigned to the head $[v_1]$ of $\sigma_1$, while its inverse $-r$ is assigned to the tail $[v_0]$. Similarly, if $\sigma^2 = [v_0,v_1,v_2]$, then $\partial_2(r\sigma^2)$ is the edge signal $r[v_1,v_2]+(-r)[v_0,v_2]+r[v_0,v_1]$. The edge $[v_0,v_2]$ reverses the boundary orientation of $\sigma^2$, hence its signal has a $-1$.

Intuitively, for the oriented simplex $[v_{i_0},\ldots, v_{i_k}]$, its boundary is just a weighted sum of of its $(k-1)$-faces. One may verify directly that $\partial_{k}\circ \partial_{k+1}=0$ \cite{Hat02}, i.e., the boundary of a boundary is $0$. This implies, for example, the image $\ima\partial_{k+1}$ of $\partial_{k+1}$ is a subspace of the kernel $\ker\partial_k$ of $\partial_k$. For each $k$, as $\partial_k$ is a linear transformation, it can be represented by a matrix $\bB_k$ w.r.t.\ bases $\{\sigma_1^k,\ldots, \sigma^k_{n_k}\}$ of $C_k(X,\bbR)$ and $\{\sigma_1^{k-1},\ldots, \sigma^{k-1}_{n_{k-1}}\}$ of $C_{k-1}(X,\bbR)$ respectively. 

Suppose $X$ is a $K$-complex, by convention, we set $C_{K+1}(X,\bbR) = C_{-1}(X,\bbR) = 0$ and $\bB_0=0$ for convenience. Then the \emph{high order Hodge Laplacians} are 
\begin{align*}
\bL_k = \bB_k\T\bB_k + \bB_{k+1}\bB_{k+1}\T, k=0,\ldots, K.
\end{align*}
For example, we may summarize aspects of traditional GSP in the new language in \cref{tab:SC}.
\begin{table}[!htb]
\caption{GSP concepts in view of the theory of simplicial complexes.}\label{tab:SC}
\centering
\scalebox{1.04}{
\begin{tabular}{ |c|c|c|c| } 
 \hline
 Graph $G$ &  Graph signals & Graph Laplacian & Connected graph\\ 
 \hline
 $1$-complex  & $C_0(G,\bbR)$ & $\bL_0$ & $\dim \ker \bL_0= 1$ \\
 \hline
\end{tabular}}
\end{table}

Compared with GSP, the theory of simplicial complexes has additional features such as the \emph{Hodge decomposition}. More precisely, it decomposes $C_k(X,\bbR)$ as the direct sum 
\begin{align} \label{eq:iok}
\ima\bB_k\T \oplus \ker \bL_k \oplus \ima\bB_{k+1}.
\end{align}
In the decomposition, $\ker (\bL_k)$ contains ``harmonic'' signals. If $k=1$, the other components are related to ``curl'' and ``divergence'' \cite{BarSar:J20}. In terms of concrete signals, for each $\bx^k \in C_k(X,\bbR)$, there exists $\bx^{k-1} \in C_{k-1}(X,\bbR)$, $\bx_H^k$ (harmonic element in $\ker (\bL_k)$) and $\bx^{k+1} \in C_{k+1}(X,\bbR)$ such that:
        \begin{align*}
            \bx^k = \bB_k\T \bx^{k-1} + \bx_H^k + \bB_{k+1}\bx^{k+1}. 
        \end{align*}
In many applications (e.g., detection of nodes emitting irregular traffics), to analyze $\bx$, we study each component of the decomposition \cite{BarSar:J20}. 

\subsection{Caution about orientation} \label{sec:cao}
\begin{figure}
    \centering
    \includegraphics[scale=0.7]{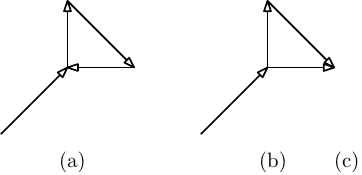}
    \includegraphics[scale=0.2]{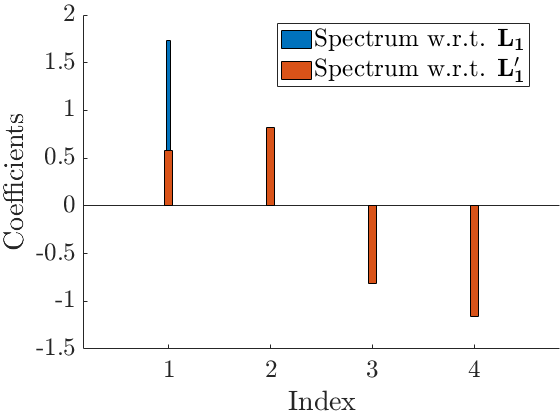}
    \caption{If we reverse the orientation of one edge in (a), we obtain the directed graph in (b) whose Hodge Laplacian $\bL_1'$ does not commute with $\bL_1$ of the graph in (a), i.e., they have different eigenvectors. We show the spectral plots of the edge signal $\bx = (0,1,1,1)\T$ w.r.t.\ $\bL_1$ and $\bL_1'$ in (c). Its spectrum is concentrated for $\bL_1$ while spread out for $\bL_1'$.}
    \label{fig:iwr}
\end{figure}

As we have seen, to define the boundary maps $\partial_k$, we need to give orientations to the simplexes. Therefore, the explicit expressions for $\bB_k$ and $\bL_k$ both depend on the orientation. In general, two different choices of orientation can result in non-commuting Hodge Laplacians $\bL_k$ and $\bL_k'$, i.e., $\bL_k'\bL_k'\neq \bL_k'\bL_k$ (\figref{fig:iwr}). Thus, the spaces of polynomial filters in $\bL_k$ and $\bL_k'$ respectively, used as filter banks in learning problems, are different. One should be careful when using Hodge Laplacians for tasks such as neural network design and interpreting signal smoothness. Procedures such as Hodge decomposition also depend on the orientation. If signals are inherently associated with orientations (e.g., flow), it is crucial that simplexes are given the correct orientations that are consistent with signal orientations. 

However, topological invariants such as Betti numbers are independent of the orientation, since spaces such as $\ker \partial_k, \ima \partial_k, \ker \bL_k$ are unique up to isomorphism.  

\section{The algebraic theory of generalized signals}\label{sec:alg_theory}

In the above setup, each signal space $C_k(X,\bbR)$ is a finite-dimensional $\bbR$-vector space, i.e., a signal is a real number. On the other hand, in the language of algebraic topology, we say that the \emph{coefficient} is $\bbR$. However, other coefficients such as integers $\mathbb{Z}$ are more often considered. On the other hand, in \cite{Ji19}, functions as signals on vertices are studied extensively, and the resulting theory is called generalized graph signal processing (GGSP). In this paper, we study such a generalized notion of signals, i.e., coefficients, based on ideas in algebraic topology, optimization, and commutative algebra \cite{Lan02}. 

\subsection{Basics of abelian groups} \label{sec:bob}
We study signals that are elements in \emph{abelian groups}. 

\begin{Definition}
    (\cite{Lan02} Chapter I) An \emph{abelian group} $A$ is a set with a commutative and associative binary operation $+$ (called the sum or addition), such that the following holds:
    \begin{enumerate}[(a)]
        \item There is the identity $0$ such that $0+a=a+0=a$ for any $a\in A$.
        \item For each $a\in A$, there is the inverse $-a \in A$ such that $a+(-a) = (-a)+a =0$.
    \end{enumerate}
        For any $n\geq 0$ and $a\in A$, $na$ is the sum of $n$-copies of $a$.
\end{Definition}

\begin{Example} \label{eg:feo}
\begin{enumerate}[(a)]
\item 
Familiar examples of abelian groups include $\mathbb{Z}$, $\bbR$, and $\mathbb{C}$ with the usual addition. For signal processing, $\bbR$ and $\mathbb{C}$ are commonly studied in Fourier theory and GSP. The space $\mathbb{Z}$ can be viewed as quantized signals (cf.\ Appendix~\ref{sec:rqs}). As a classical application, considering signals in $\mathbb{Z}$ can be used for space classification, as we shall see in \cref{eg:trp}.

Function spaces such as polynomials $\bbR[t]$, splines, or $L^2(\Omega)$ over a domain $\Omega$ are also abelian groups, with the usual addition of functions. They are used as signals for continuous time-vertex graph signal processing \cite{Ji19}. For example, such a signal can be used to model temporal correlations of time series. As we have exemplified in \cref{sec:intro}, it can serve as an alternative to ordinary signals in $\mathbb{R}$, in situations such as asynchronous sampling.

We also have the finite abelian group $\mathbb{Z}/n\mathbb{Z}$ consisting of elements $\{0,\ldots,n-1\}$, with addition given by addition with the modulo operation $\mathrm{mod}\ n$. The case $n=2$ corresponds to $0,1$ bits and is related to coding theory (cf.\ \cref{eg:ika} below).

%\ifarxiv
\item \underline{Case study: differential forms}. We describe the main idea regarding an ``exotic'' but important example in calculus. Details can be found in \cite{Rud76} p.\ 245. 

Let $A$ be the group of \emph{differential $k$-forms} on the standard $k$-simplex $\Delta_k$. An element $\omega\in A$ takes the form 
\begin{align*}
\omega = f \ud x_0\wedge\ldots\wedge\ud x_k,
\end{align*}
where $f$ is a smooth function on $\Delta_k$, and $\ud x_0\wedge\ldots\wedge\ud x_k$ is a formal symbol that corresponds to the infinitesimal volume $\ud x_0\ldots\ud x_k$ for integration. The addition on $A$ is induced by the function addition of the leading term $f$ and $A$ is an abelian group. We define $\int_{\Delta_k}\omega$ by integrating $f$ over $\Delta_k$ using the Riemann or Lebesgue integral. 

Consider a ``nice'' simplicial complex $X$ (cf.\ \cite[Section 10.26]{Rud76}) of dimension $k$. For a signal $\bomega$ in $C_k(X,A)$ that assigns a differential form to each $k$-simplex of $X$, we can extend the integral to obtain $\int_X \bomega$, by taking the sum of the integrals over the $k$-simplexes in $X$. On the other hand, if $\bomega'$ is a $(k-1)$-form on $X$, then taking partial derivatives and applying the chain rule (in multivariate calculus) results in a $k$-form $\ud\bomega' \in C_k(X,A)$. 
 
The setup in terms of $C_k(X,A)$ permits one to formulate and show important results such as Stoke's theorem: 
\begin{align*}
    \int_X \ud\bomega' = \int_{\partial_k X} \bomega'.
\end{align*}
%\fi

\end{enumerate}
\end{Example} 

We next describe a few useful constructions that build new groups from old ones. Given an abelian group $A$, a subset $A'$ is a \emph{subgroup} if $A'$ is also an abelian group, i.e., $0\in A'$ and if $a$ is in $A'$, then so is $-a$. We may form the \emph{quotient group} $A/A'$. As a set, each element of $A/A'$ is a class of the form $[a]:= a+A'=\set{a+a'\given a'\in A'}$ for $a\in A$. The addition on $A/A'$ is defined by $[a]+[b] = [a+b]$. For example, the notation $\mathbb{Z}/n\mathbb{Z}$ (in \cref{eg:feo}) suggests that it can also be interpreted as a quotient group. In $\mathbb{Z}$, $n$ multiples of integers form a subgroup $n\mathbb{Z}$. Then as a quotient group, each element of $\mathbb{Z}/n\mathbb{Z}$ is the class of the form $[a] = \{a+nb \mid b\in \mathbb{Z}\}$. We use the modulo operation because $[a+b] = [a+b\mod n]$ as sets and hence each class has a unique representative among $\{0,\ldots,n-1\}$.

Another useful construction is the direct sum. Given abelian groups $A$ and $B$, their \emph{direct sum} $A\oplus B$ consists of pairs $(a,b)$ with addition $(a,b)+(a',b') = (a+a',b+b')$. This is analogous to the direct sum of vector spaces. In particular, for $n>0$, $A^n$ is the direct sum of $n$-copies of $A$. 

A \emph{homomorphism} $\phi: A \to A'$ between abelian groups $A$ and $A'$ is a function that preserves the addition, i.e., $\phi(a+b) = \phi(a)+\phi(b)$. It is an \emph{isomorphism} if $\phi$ is a bijection and we say that $A$ and $A'$ are \emph{isomorphic}, denoted by $A\cong A'$. Usually, isomorphic abelian groups are regarded as the same algebraic object. The \emph{kernel} $\ker\phi$ of $\phi$ is the inverse image of $0$ in $A'$, i.e., $\phi^{-1}(0)$. It is a subgroup of $A$. On the other hand, the image $\ima\phi$ of $\phi$ is a subgroup of $A'$. The (first) isomorphism theorem \cite{Lan02} claims that $A/\ker\phi \cong \ima\phi$.

\subsection{A brief introduction to rings and modules} \label{sec:abi}
An abelian group $A$ usually carries additional algebraic structures (e.g., a ring or a module) that can be useful. In this section, we formally define rings and modules, with prototypical examples $\bbR$ and $\bbR^n$. Though they are not essential in many discussions, knowing these concepts can be helpful to better understand the discussions. Details can be found in textbooks such as \cite{Lan02,Hun74}.

A \emph{commutative ring} $R$ is a set with two binary operations $+,\cdot$ such that the following holds
\begin{enumerate}[(a)]
    \item $R$ is an abelian group with $+$ and the identity $0$. 
    \item The multiplication $\cdot$ on $R$ is commutative and associative. By convention, we may omit the multiplication symbol if no confusion arises.
    \item There is a multiplicative identity $1$ such that $1\cdot r = r\cdot 1 = r$ for any $r\in R$.
    \item The multiplication $\cdot$ is distributive w.r.t.\ $+$.
\end{enumerate}

For example, $\mathbb{Z}$, $\bbR$, and $\mathbb{C}$ are all rings with the usual addition and multiplication. However, unlike $\bbR$ and $\mathbb{C}$, for each nonzero $n\in \mathbb{Z}$, its inverse $n^{-1}$ is not contained in $\mathbb{Z}$ unless $n=\pm 1$. Therefore, in a ring, it is not required that a nonzero element has a multiplicative inverse. Function groups such as $\bbR[t]$ and $L^2(\Omega)$ are rings with the pointwise multiplication of functions. The group $\mathbb{Z}/n\mathbb{Z}$ is also a ring, and the multiplication is defined by applying the modulo $n$ operation to the usual multiplication.

Given a ring $R$, an abelian group $M$ is a \emph{module over $R$} if there is a scalar multiplication $\cdot: R\times M \to M$ (the symbol ``$\cdot$'' is often omitted) such that the following holds for $r,r'\in R$ and $x,y\in M$:
\begin{enumerate}[(a)]
    \item $r\cdot (x+y) = r\cdot x + r\cdot y$; $(r+r')\cdot x = r\cdot x + r'\cdot x$.
    \item $(rr')\cdot x = r\cdot (r'\cdot x)$.
    \item $1\cdot x = x$.
\end{enumerate}

The most familiar example of a module is a finite-dimensional vector space, with $\bbR^n$ being a module over the ring of real numbers $\bbR$. Each ring $R$ is a module over itself, with scalar multiplication the same as the ring multiplication.

\subsection{Signals on simplicial complexes}\label{sec:SSC}

With enough preparation, the algebraic theory of generalized signals, mutatis mutandis, can be developed analogously to \cref{sec:top}. We follow \cref{sec:top}, while also highlighting essential differences. We use the same notations regarding simplicial complexes.

For an abelian group $A$, the \emph{$k$-chain $C_k(X,A)$ with coefficient $A$} is $A^{n_k}$. For each $1\leq i\leq n_k$, the $i$-component of $A^{n_k}$ can be interpreted as signals (in $A$) on the oriented $k$-simplex $\sigma^k_i$. Similar to $C_k(X,\bbR)$, an element $(a_1,\ldots, a_{n_k})$ in $C_k(X,A)$ can be identified with a formal sum $\sum_{1\leq i\leq n_k}a_k\sigma^k_i$, with each $\sigma^k_i$ treated as a basis element. We usually do not distinguish these two interpretations of $C_k(X,A)$. If $A$ is a module over a ring $R$, then $A^{n_k}$ is also a module over $R$ with the scalar multiplication $r(a_1,\ldots, a_{n_k}) = (ra_1,\ldots,ra_{n_k})$. Therefore, in this paper, $C_k(X,A)$ is a module over $R$.

The boundary map $\partial_k$ can be defined by the exact same formula as in \cref{defn:ssi}. \emph{Notational convention}: we may also write $\partial_{k,A}$ or $\partial_{k,A,X}$ on rare occasions to highlight the group $A$ and the simplicial complex $X$. 

By stacking $\partial_k$ for all $k$ together, we obtain a sequence of boundary maps:
\begin{align} \label{eq:cxp}
\begin{split}
\cdots \xrightarrow{\partial_{k+1}} &\; C_k(X,A) \xrightarrow{\partial_k} C_{k-1}(X,A) \xrightarrow{\partial_{k-1}} \\ \cdots & \xrightarrow{\partial_2} C_1(X,A) \xrightarrow{\partial_1} C_0(X,A) \to 0.
\end{split}
\end{align}
We still have the key identity $\partial_k\circ \partial_{k+1}=0$, and the sequence in (\ref{eq:cxp}) is called a \emph{chain complex} \cite{Hat02}. The upshot of the identity is that $\ima\partial_{k+1}$ is a subgroup of $\ker\partial_k$ and the quotient group plays an important role \cite{Hat02}.

\begin{Definition} \label{defn:thh}
    The $k$-th homology group of $X$ with coefficient $A$ is $H_k(X,A) = \ker\partial_k/\ima\partial_{k+1}$.
\end{Definition}

As introduced in \cref{sec:bob} for quotient group, each element of $H_k(X,A)$ is a class $[\bx] = \{\bx+\by \mid \by \in \ima \partial_{k+1} \}$ with $\partial_k(\bx)=0$. The homology group can be compared to something we are more familiar with. 

\begin{Theorem}\label{lem:fam}
If $A$ is an $\bbR$-vector space, we have $H_k(X,A) \cong \ker\bL_k$. For a general $A$ that is \emph{torsion free} (i.e., $na = 0$ with $a\in A, n\in \mathbb{Z}$ implies either $n=0$ or $a=0$), we have $\ker\bL_k \subset H_k(X,A)$.
\end{Theorem}

The boundary maps $\partial_k$ generate signals on lower dimensional simplexes from those on high dimensional simplexes. We can reverse the process by taking the transpose $\partial_k^*: C_{k-1}(X,A) \to C_k(X,A)$ of $\partial_k$. Similar to (\ref{eq:cxp}), we also have a sequence of maps, called the \emph{cochain complex}, in the reverse direction:
\begin{align} \label{eq:cxc}
\begin{split}
    \cdots \xleftarrow{\partial_{k+1}^*} & C_k(X,A) \xleftarrow{\partial_k^*} C_{k-1}(X,A) \xleftarrow{\partial_{k-1}^*} \\ \cdots & \xleftarrow{\partial_2^*} C_1(X,A) \xleftarrow{\partial_1^*} C_0(X,A) \xleftarrow{} 0.
    \end{split}
\end{align}

Mimicking \cref{defn:thh}, we can define the $k$-th \emph{cohomology} of $X$ as $H^k(X,A) = \ker \partial_{k+1}^* / \ima \partial_k^*$. In our setting that $X$ being finite, $H_k(X,\mathbb{R}) = H^k(X,\mathbb{R})$ (\cite{Hat02} p.195), and we shall mainly focus on the homology groups. 

In view of \cref{lem:fam}, if $A = \bbR$, the Hodge decomposition (\ref{eq:iok}) is essentially the isomorphic identity 
\begin{align} \label{eg:cbc}
C_k(X,\bbR) \cong \ima\partial_k^* \oplus H_k(X,\bbR) \oplus \ima\partial_{k+1}.
\end{align}

However, though useful, for a general $A$ such as $\mathbb{Z}$, we do not always have a Hodge decomposition (\ref{eq:iok}) as we demonstrate with the next example.

\begin{Example} \label{eg:trp}
\underline{Case study: classifying topological spaces.} 
Real projective planes are mathematical models of the optical effects we see when parallel lines seem to converge \cite{Lam17}. The study of geometry on such a plane, aka projective geometry, dates back to the Renaissance when artists started to use perspective in paintings. 

A compact model $X$ for real projective planes is rigorously defined as the quotient space:
    \begin{align*}
    [0,1]\times [0,1]/\{(0,y)\sim (1,1-y), (x,0)\sim (1-x,1)\}, 
    \end{align*}
i.e., $X$ can be obtained by gluing opposite sides of a square in reverse directions. $X$ has two $2$-simplexes $\sigma_1^2, \sigma_2^2$, three edges $\sigma_1^1, \sigma_2^1, \sigma_3^1$ and two nodes $v_1, v_2$, as shown in \figref{fig:rp2}(a). If $A = \mathbb{Z}$, it can be computed (see \cite{Hat02} for details) that $H_1(X,A) \cong \mathbb{Z}/2\mathbb{Z}$ and $\ker \bL_1 = 0$. Notice that $H_1(X,\mathbb{Z})$ is finite, while $C_1(X,\mathbb{Z}) = \mathbb{Z}^3$ does not have any nontrivial finite subgroup. Hence, the Hodge decomposition (\ref{eg:cbc}) does not hold. 
\begin{figure}[h]
    \centering
\includegraphics[width=0.8\columnwidth]{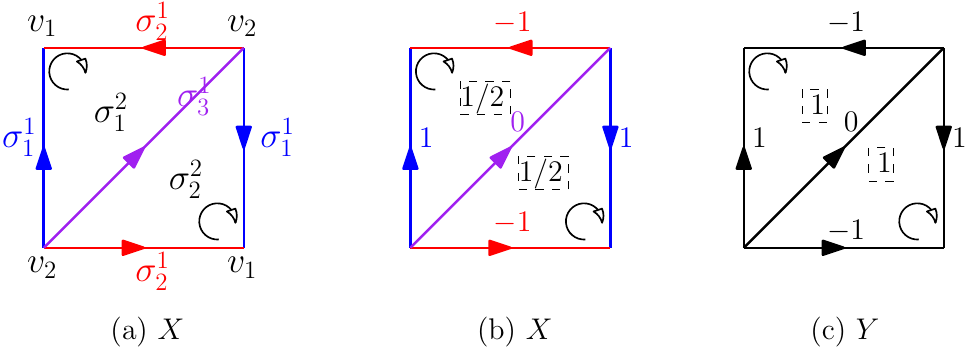}
    \caption{The (oriented) simplicial structure of a real projective plane model $X$ is shown in (a). Opposite sides are identified according to the shown directions. In (b) and (c), we show edge signals that distinguish $X$ from a model $Y$ of the ordinary plane. In dashed squares, we show the triangle signals. }
    \label{fig:rp2}
\end{figure}

We may use edge signals on $X$ to distinguish it from a (ordinary) plane. The latter is topologically equivalent to an ordinary square $Y$. It has a similar triangulation (\figref{fig:rp2}(c)), but without identifying opposite sides. We construct $\bx \in C_1(X,\bbZ)$ and $\by \in C_1(Y,\bbZ)$ by assigning edge signals as in \figref{fig:rp2}(b) and (c). By \cref{defn:ssi} and applying the follow-up examples, we see that $\bx\in\ker \partial_{1,\mathbb{Z},X}$ and $\by\in\ker \partial_{1,\mathbb{Z},Y}$ respectively (e.g., the signal at node $v_1$ in $X$ takes $1$ as contribution from $\partial_{1,\mathbb{Z},X} (\sigma^1_1)$ and $-1$ from $\partial_{1,\mathbb{Z},X} (\sigma^1_2)$). However, $\bx \notin \ima \partial_{2,\mathbb{Z},X}$ since $\bx = \partial_{2,\mathbb{Z},X}\big((1/2,1/2)\T\big)$ and $1/2 \notin \mathbb{Z}$, while $\by = \partial_{2,\mathbb{Z},Y}\big((1,1)\T\big) \in \ima \partial_{2,\mathbb{Z},Y}$. The difference is due to that in $X$, the top and bottom sides (resp. left and right sides) are both identified as $\sigma_2^1$ (resp.\ $\sigma_1^1$) and the corresponding component of $\ima \partial_{2,\mathbb{Z},X}$ takes contribution from both triangles $\sigma_1^2$ and $\sigma_2^2$. By a fundamental result in algebraic topology, $X$ and $Y$ are not topologically equivalent due to $\bx \notin \ima \partial_{2,\mathbb{Z},X}$. 

The argument does not work if we consider real signals $A =\bbR$, as $1/2 \in \bbR$ and $\bx \in \partial_{2,\mathbb{R},X}$. Therefore, considering integer signals gives us additional information about the spaces, and this is one of the primary reasons that integers are usually considered in favor of real numbers in space classification. On the surface, $A = \mathbb{Z}$ benefits from having non-invertible elements, however, there are deeper algebraic reasons related to the fact that $\mathbb{Z}$ has prime numbers \cite{Lan02}, which we will not explore further here.    
\end{Example}
%https://www.ias.edu/video/quantum-error-correction-systolic-geometry-and-probabilistic-embeddings

For a general $A$, we also want to study a signal $\bx \in C_k(X,A)$ by considering three signals $\bx_1 \in \ima\partial_{k+1}$, $\bx_{-1} \in \ima\partial_k^*$ and $\bx_0$ associated with $H_k(X,A)$ such that they collectively give as much information as contained in $\bx$. One challenge is that we do not have a unique algebraic decomposition $\bx = \bx_{-1} + \bx_0 + \bx_1$, since each element of the quotient group $H_k(X,A)$ is a class. Therefore, we propose a learning framework for this purpose in the next section.

\section{The Learning Framework}\label{sec:learning}

In this section, we develop an approach based on optimization to substitute some algebraic constructions, such as the Hodge decomposition, for general coefficient groups $A$.

To start with, assume $A$ has a metric $d(\cdot,\cdot)$ that satisfies $d(a,a')=d(a+b,a'+b)$ for any $a,a',b\in A$. For any $a\in A$, let its \emph{norm} be $\norm{a}_A = d(a,0)$. It can be verified that $\norm{a+a'}_A\leq \norm{a}_A+\norm{a'}_A, a,a'\in A$. 

Moreover, if $A$ is a module over a ring $R$ with norm $\norm{\cdot}_R$, then $\norm{\cdot}_A$ is assumed to satisfy $\norm{ra}_A= \norm{r}_R\norm{a}_A$, for all $a\in A, r\in R$. We mostly consider only $R=\mathbb{Z},\bbR,\mathbb{C}$, or $R=A$ where $A$ itself is a ring.
 
For example, if $A = \mathbb{Z}, \bbR, \mathbb{C}$, we use the Euclidean metric for $d(\cdot,\cdot)$. For $A = L^2(\Omega)$ and its subgroups, we use the $L^2$-metric. For $A = \mathbb{Z}/2\mathbb{Z}$, the discrete metric is a natural choice.

The metrical setup can be extended to $C_k(X,A)$. Recall that the standard basis of $C_k(X,A)$ is $\{\sigma^k_1,\ldots, \sigma^k_{n_k}\}$, with each $\sigma^k_i$ an oriented $k$-simplex of $X$ and $n_k$ is the number of $k$-simplexes. 

\begin{Definition}
    Let $\bw = (w_i)_{1\leq i\leq n_k}$ be a set of non-negative weights of $k$-simplexes. By default, we let the components of $\bw$ be all $1$s if unspecified. Then for $\bx = \sum_{1\leq i\leq n_k} a_i\sigma^k_i \in C_k(X,A)$ where each $a_i\in R$, its $\bw$-weighted \emph{$p$-norm} is 
    \begin{align*}
        \norm{\bx}_{X,k,p}^p = \sum_{1\leq i\leq n_k} w_i^p\norm{a_i}_A^p.
    \end{align*}
    For convenience, we may suppress $X$ in the subscript as $\norm{\cdot}_{k,p}$ if no confusion arises.
\end{Definition}

We usually consider $p=2$ for $A = \bbR, \mathbb{C}, L^2(\Omega)$. On the other hand, from \cref{eg:ika} below, other choices of $p$ can also be interesting. Moreover, The norm can be extended to the quotient group $H_k(X,A)$ of $ C_k(X,A)$.  

\begin{Definition} \label{defn:fab}
The simplicial (semi) norm of $\by \in H_k(X,A)$ is
\begin{align*}
    \norm{\by}_p = \inf_{[\bx]=\by} \norm{\bx}_{k,p},
\end{align*}
where the infimum is taken over all $\bx\in \ker\partial_k \subset C_k(X,A)$ whose class in $H_k(X,A)$ is $\by$. 
\end{Definition}

\begin{Example} \label{eg:ika}
\begin{enumerate}[(a)]
    \item \red{[For arXiv only.]} If $k=0$ and $X=G$ is a connected graph, then any $\bx \in C_0(X,\bbR)$ is an ordinary graph signal. Let $\by = [\bx]$ be the equivalence class of $\bx$ in $H_0(G,\bbR) \cong \bbR$. It can be checked that $\bx'\in C_0(X,\bbR)$ satisfies $[\bx']=\by$ if and only if $\bx'$ and $\bx$ have the same component sums. From this, we obtain that $\norm{\by}_2$ is realized by the constant signal, with each component the average of $\bx$. This is nothing but the projection of $\bx$ to the space of constant signals as in classical GSP.     
    \item \label{it:csc} 
    \underline{Case study: coding.} We demonstrate how coefficients other than $\mathbb{Z},\mathbb{R}$ and functions may arise. Moreover, we consider $p=1$ instead of $2$ for the norm. $A = \mathbb{Z}/2\mathbb{Z}$ is used to construct linear codes such as expander, Tanner, and homology codes \cite{Sip96,Etz99,Bom07}. Following \cite{Bom07}, for a graph $G$, codewords form the subspace of edge signals $\ker \partial_1$ in $C_1(G,A)$. The matrix $\bB_1$ of $\partial_1$ forms a parity check matrix. 

    For a prototypical example, consider $G=C_n$ the cycle graph of size $n$. As $-1=1$ in $A = \mathbb{Z}/2\mathbb{Z}$, we have 
    \begin{equation*}
    \bB_1 =
    \begin{pmatrix}
        1 & 0 & 0 & \ldots & 0 & 1 \\
        1 & 1 & 0 & \ldots & 0 & 0 \\
        0 & 1 & 1 & \ldots & 0 & 0 \\
        \ldots & & & & & \ldots \\
        0 & 0 & 0 & \ldots & 1 & 0 \\
        0 & 0 & 0 & \ldots & 1 & 1
    \end{pmatrix}.
    \end{equation*}
    There are only two signals, and hence codewords, in $\ker \partial_1$: $(0 0 \ldots 0)\T$ and $(1 1 \ldots 1)\T$. Therefore, the code associated with $C_n$ is nothing but the \emph{$n$-bit repetition code}. For parity check, each row of $\bB_1$ sums up bits on a pair of adjacent edges of $G$. A signal $\bx=(a_1, a_2, \ldots, a_n)\T$ satisfies $\bB_1\bx = 0$ if and only if $a_1=-a_n=a_n$ and $a_i=-a_{i+1}=a_{i+1}, 1\leq i\leq n-1$, i.e., $\bx$ is one of the two codewords. 
    
    In general, invariants of the code, such as code length, dimension, and distance, can be estimated and analyzed using geometric invariants of $G$ (cf.\ \cite{Bom07} Theorem II.2.). If $p=1$, the $1$-norm with default $\bw$ is nothing but the Hamming distance. The smallest non-zero $\norm{\by}_p$ for $0\neq \by \in H_1(G,A)$ measures the smallest size of a ``cycle'' in $G$ \cite{Kat07}, which is closely related to the distance of the code. 
    
    In the example of the $n$-bit repetition code above, $G =C_n$ has $n$ edges, and $n$ is the \emph{length} of the code, defined as the dimension of the ambient code space. $G$ has a single cycle, which corresponds to the \emph{dimension} of the code. The cycle length is $n$. By no coincidence, it is the \emph{code distance}, defined as the shortest Hamming distance of a nonzero codeword, of the repetition code.  
    
    Correspondingly, there is a quantum version for a general simplicial complex $X$ \cite{Cal96, Ste96, Bom07, Bro14}. While impossible to give the details here due to space constraints, we indicate the group $A$ involved.  Instead of using $A = \mathbb{Z}/2\mathbb{Z}$ for a bit, a physical qubit belongs to $A = \mathbb{C}^2$. The pair of standard basis vectors of $A$ correspond to $0$ and $1$ in a classical bit. For example, a toric code assigns a physical qubit to an edge on the lattice of a torus (\figref{fig:atc}). On the other hand, a type of error operator can be identified with $C_k(X,A')$, where $A'$ is generated by the Pauli $Z$-operator \cite{Bro14} 1.2.2. Optimal decoding corresponds to finding the simplicial norm of a homology class. Combinatorial methods such as minimal weight perfect matching \cite{Bro14} 2.2.7 can be applied to the minimization in \cref{defn:fab}.    
\begin{figure}
    \centering
    \includegraphics[scale = 0.45]{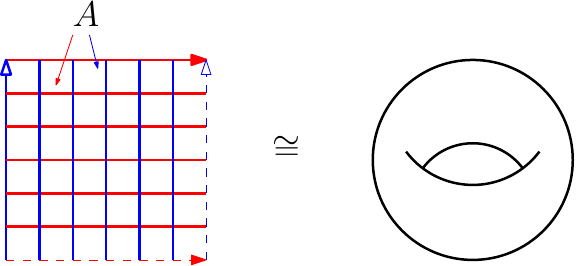}
    \caption{A torus can be obtained by identifying pairs of sides of a square in the same direction. Thus, a lattice graph $L$ on a torus can be visualized as a lattice on the square. This cellulation is analogous to a triangulation of the torus using squares instead of triangles. For a toric quantum code \cite{Bro14}, a codeword is a signal in $C_1(L,A)$ with $A = \mathbb{C}^2$.}
    \label{fig:atc}
\end{figure}
    
\end{enumerate}
\end{Example}

As discussed in \cref{sec:SSC}, given a signal $\bx \in C_k(X,A)$ for a general $A$, we usually do not have the Hodge decomposition. 
Inspired by \cref{lem:fam}, we replace the harmonic component $\ker \bL_k$ by $H_k(X,A)$. As each element of $H_k(X,A)$ is a class in $\ker \partial_k$, we introduce an optimization analogous to \cref{defn:fab} to pinpoint a single element. Hence, we propose a fundamental learning model as the following optimization: 
\begin{align}\label{eq:mbb}
\begin{aligned}
    \min_{(\bx_0 ,\bx_1,\bx_{-1})} &\norm{\bx_0}_{k,p} + \zeta_1 \norm{\bx'-\bx}_{k,p} + \zeta_2 R(\cdot) \\
    \ST & \bx_0 + \bx_1 + \bx_{-1} = \bx'\\
    & \bx_0 \in \ker\partial_k,\\
    & \bx_1 = \partial_{k+1}(\by_1),\ \by_1\in C_{k+1}(X,A),\\
    & \bx_{-1} = \partial_k^*(\by_{-1}),\ \by_{-1}\in C_{k-1}(X,A).
\end{aligned}
\end{align}
The term $\norm{\bx_0}_{k,p}$ corresponds to the Hodge decomposition (cf.\ \cref{thm:fam}) where $A=\bbR$ and $p=2$, while $\norm{\bx'-\bx}_{k,p}$ is the term for data fidelity. Lastly, $R(\cdot)$ is a regularizer that is problem-dependent. It usually reflects our prior knowledge of the properties of the signal such as smoothness as well as task-specific signal models. The coefficients $\zeta_1,\zeta_2$ are tunable hyperparameters. The analytic method based on optimization presented here can be applied for a general $A$, while the algebraic way has a strong restriction on $A$. 

\begin{Theorem} \label{thm:fam}
For $A = \bbR$ and $p=2$, if the parameter $\zeta_1>1, \zeta_2=0$ in (\ref{eq:mbb}), then the Hodge decomposition of $\bx'=\bx$ is the unique solution to the fundamental learning model (\ref{eq:mbb}). 
\end{Theorem}

There are a few benefits to using (\ref{eq:mbb}):
\begin{enumerate}[(a)]
    \item Primarily, it can be used as a theoretical substitute for the Hodge decomposition when the latter is not applicable.
    \item It is more flexible. In practice, problem (\ref{eq:mbb}) can be readily modified for different applications. For example, if $\bx$ is only partially observed (in some components) as $\bz$, the fidelity term $\zeta_1\norm{\bx' - \bx}_{k,p}$ can be replaced by $\zeta_1\norm{\bP(\bx')-\bz}_{k,p}$, where $\bP$ is the projection operator to the coordinate space of the observed signals. We shall see other variants of such a modification when we discuss filters in \cref{sec:fil}.
    \item The metrical approach may offer a different perspective. For example, consider $A$ the ring of degree $k$ polynomials. If we view $A$ as a vector space spanned by $\{1,t,\ldots, t^k\}$, the ordinary Hodge decomposition of each component is essentially solving (\ref{eq:mbb}) with $A$ given the Euclidean metric on the vector of coefficients. 
    However, for function spaces, it is more common to consider the $L^2$-norm, which is independent of the choice of function basis. Moreover, it distinguishes the functions better. For example, the basis monomials have different $L^2$-norms, while they all have norm $1$ when using the above Euclidean metric.
    Hence, the proposed learning model can be more useful. For another example, if $A$ is the group of cubic splines, then $L^2$-norm of the $2$nd order derivative is usually used.
\end{enumerate}

In practice, to solve (\ref{eq:mbb}), we need to find generators of $C_{k-1}(X,A)$, $C_{k+1}(X,A)$ and $\ker\partial_k$ so that the problem is parametrized. Both $C_{k-1}(X,A)$ and $C_{k+1}(X,A)$ already have the canonical generators $\set{\sigma_i^{k-1}\given i=1,\dots,n_{k-1}}, \set{\sigma_i^{k+1}\given i=1,\dots,n_{k+1}}$. For $\ker\partial_k$, the following version of the \emph{universal coefficient theorem}, also used in proving \cref{lem:fam}, can be useful. Recall that $A$ is said to be torsion-free if $na = 0$ with $a\in A, n\in \mathbb{Z}$ implies either $n=0$ or $a=0$.

\begin{Theorem}[\cite{Hat02} p.264] \label{thm:iai}
 If $A$ is torsion free, then $H_i(X,A)\cong H_i(X,\mathbb{Z})\otimes A$, where $\otimes$ is the tensor product. Moreover, if $A$ is an $\bbR$-vector space, then $H_i(X,A)\cong H_i(X,\bbR)\otimes A$.   
\end{Theorem}

All of $\bbR, \mathbb{C}, \bbR[t], L^2(\Omega)$ and $\mathbb{Z}/2\mathbb{Z}$ are torsion-free. As suggested by the result, we may apply the following framework to parameterize $\ker \partial_k$ in (\ref{eq:mbb}). Suppose $\{\bx_1,\ldots,\bx_m\}$ is a set of generators of $\ker \partial_{k,\mathbb{Z}}$. Then each element $\by$ of $\ker \partial_{k,A}$ can be expressed as a combination $\sum_{1\leq i\leq m} a_i\bx_m$ with $a_i\in A$. In an optimization, one estimates each coefficient $a_i$.

In Appendix~\ref{sec:rqs}, we focus on the optimization aspect of quantized signals when $A=\mathbb{Z}$, illustrated with a study on a co-authorship complex \cite{Ebl20}.

Though not used in the sequel, we mention that the fundamental learning framework can approximate algebraic procedures such as band-pass filters w.r.t.\ the Hodge Laplacian. Details are provided in Appendix~\ref{sec:abf}. 

\section{Filters} \label{sec:fil}
Filters play an essential role in signal processing that allows us to capture relations between signals. In this section, we introduce the notion of filters for general coefficient groups. 

For simplicial complexes $X,X'$ and abelian groups $A,A'$, let $\bF: C_k(X,A) \to C_k(X',A')$ be a homomorphism of abelian groups. It is called a \emph{filter} if the following holds:
\begin{itemize}
    \item $\bF(\ker\partial_{k,A,X}) \subset \ker\partial_{k,A',X'}$;
    \item $\bF(\ima \partial_{k+1,A,X}) \subset \ima \partial_{k+1,A',X'}$;
    \item $\bF(\ima \partial_{k,A,X}^*) \subset \ima \partial_{k,A',X'}^*$. 
\end{itemize}
Intuitively, $\bF$ preserves all the subgroups of $C_k(\cdot,\cdot)$ we are interested in. Suppose $X=X'$ and $A=A'$. If $S$ is a subgroup of $C_k(X,A)$, a filter $\bF$ is called \emph{$S$-bandlimited} if $\bF(S) \subset S$. 

For common application scenario (e.g., \cref{sec:hfb}), it is often true that $\bF$ is chosen from a linearly parameterized family, i.e., there are fixed (known) filters $\bF_1,\ldots, \bF_l$ such that $\bF = \sum_{1\leq i\leq } a_i\bF_i, a_i\in \mathbb{R}$. Then $\bF$ can be estimated using an optimization framework similar to classical GSP using training datasets. Specifically, if $\bx$ and $\bz$ are the source and target signals, then we may estimate $a_1,\ldots, a_l$ to determine $\bF$ by replacing the (middle) fidelity term in (\ref{eq:mbb}) with 
\begin{align*}
    \norm{\sum_{1\leq i\leq l} a_i\bF_i(\bx')-\bz}_{k,p}. 
\end{align*}
For a concrete example, if $A$ is a module over a ring $R$ (cf.\ \cref{sec:abi}), then analogous to GSP, a polynomial (aka convolution in GSP \cite{Shu13}) in $\bL_k$ with coefficients in $R$ is always a filter on $C_k(X,A)$ \cite{Ebl20}. In this case, to estimate such a filter, the basis is $\{\bL_k,\ldots, \bL_k^l\}$. However, as we have discussed in \cref{sec:cao}, readers should be careful when using these filters. 

We next focus on less conventional constructions of filters, as compared with GSP, which might be of interest. 

\subsection{Subdivision} \label{sec:sd}

In this subsection, we describe how filters arise from a type of topological construction. A limitation of TSP including GSP is that many concepts are not preserved under homeomorphism of simplicial complexes, needless to say, under homotopy equivalence \cite{Hat02}. For example, signal space dimensions and Laplacians depend on specific simplicial structures. For \emph{simplicial subdivision}, which is a homeomorphism that changes the simplicial structure, we propose to use filters to partially address the above shortcomings. 

Recall that a subdivision of a simplicial complex is a refinement of the simplicial structure while maintaining the topology \cite{Hat02}. More precisely, a simplicial complex $Y$ is a subdivision of $X$ if the following holds:
\begin{itemize}
    \item $|Y|=|X|$, i.e., $X, Y$ have the same geometric realization (see \cref{sec:sco}).
    \item Each simplex of $Y$ is contained in a simplex of $X$.
\end{itemize}
This means that for each simplex $K$ of $X$, its geometric realization is that of a union of $m_K$ simplexes $L_1,\ldots, L_{m_K}$ in $Y$. We assume that the orientation of $L_i, 1\leq i\leq m_K$ are given such that: 
\begin{itemize}
    \item The simplexes of $L_i$ contained in $K$ are ordered according to that of $K$.
    \item Common simplexes of $L_i, L_j$ are oriented in the opposite direction (e.g., \figref{fig:tfi}).
    \begin{figure}
        \centering
        \includegraphics[scale=0.6]{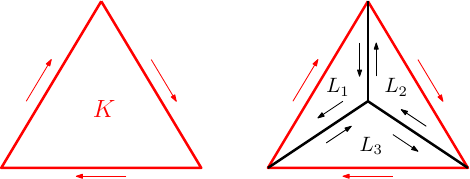}
        \caption{The figure illustrates the subdivision of a $2$-complex $K$ into $L_1,L_2$ and $L_3$ using the barycenter of $K$. Common edges of $L_1, L_2, L_3$ are oriented in opposite directions.}
        \label{fig:tfi}
    \end{figure}
\end{itemize}

If a signal assigns $x\in A$ to the simplex $K$, then we \emph{extend} it naturally to $Y$ by assigning $x$ to each $L_i,1\leq i\leq m_K$. In this way, any $\bx \in C_k(X,A)$ extends to a signal $\widetilde{\bx}$ in $C_k(Y,A)$. Regarding $\bw$ in computing norms, we assume that the weight $w$ of $\bw$ on each $K$ is partitioned into weights on $L_1,\ldots, L_{m_k}$ such that the $p$-norm of the weight is preserved. 

In additional to $|X|=|Y|$, the next property implies that $\mathcal{E}: C_k(X,A) \to C_k(Y,A), \bx \mapsto \widetilde{\bx}$ is a filter.  
%\red{[Confusing notations. Is $\widetilde{\bx}_0 = \calE(\bx)_0$ How about $\widetilde{\bx_0}$? Clearer to use $\calE$ notation instead of $\widetilde{\cdot}$.]}

\begin{Lemma} \label{lem:ibb}
If $\bx = \bx_{-1} + \bx_0 + \bx_1$ with $\bx_{-1} \in \ima\partial_k^*$, $\bx_1 \in \ima\partial_{k+1}$ and $\bx_0 \in \ker\partial_k$ (in $X$), then $\mathcal{E}(\bx) = \mathcal{E}(\bx_{-1}) + \mathcal{E}(\bx_0) + \mathcal{E}(\bx_1)$ with $\mathcal{E}(\bx_{-1}) \in \ima\partial_k^*$, $\mathcal{E}(\bx_1) \in \ima\partial_{k+1}$ and $\mathcal{E}(\bx_0) \in \ker\partial_k$ (in $Y$). Moreover, $\mathcal{E}$ is contracting, i.e.,  
\begin{align*}
\norm{\mathcal{E}(\bx)_0}_{Y,k,p} \leq \norm{\mathcal{E}(\bx_0)}_{Y,k,p} = \norm{\bx_0}_{X,k,p}.
\end{align*} 
\end{Lemma}

Subdivision allows us to perform signal processing on $Y$ with a more refined simplicial structure given a signal on $X$, though topologically $X$ and $Y$ are essentially the same. A numerical study can be found in \cref{sec:hfb}.  

\subsection{Change-of-coefficient filters and related constructions} \label{sec:coc}

If $\phi: A \to A'$ is an abelian group homomorphism (e.g., the inclusion homomorphism $A=\mathbb{Z} \to A' =\bbR$ in Appendix~\ref{sec:rqs}), then it induces a transformation $\phi_{*,k}: C_k(X,A) \to C_k(X,A')$ by applying $\phi$ to each component of $C_k(X,A)\cong A^{n_k}$. 
\begin{Lemma} \label{lem:pia}
    $\phi_{*,k}$ is a filter.
\end{Lemma}

\emph{Interpretation of shift}. Let $A=A' = L^2(\bbR)$. We have the translation homomorphism $\phi(f)(t) = f(t-T)$ for $f\in L^2(\bbR), T\neq 0$. 
If the domain $\bbR$ of the function space is considered as the time domain, then this example describes a shift in this domain. 
It is worth describing a related construction regarding such a shift, which is interesting in its own right.

The shift defined above arises from a translation $\psi: \bbR \to \bbR, t \mapsto t+T$. In general, suppose $\Omega$ is a topological space and $\psi: \Omega \to \Omega$ is a homeomorphism, i.e., a continuous function with a continuous inverse $\psi^{-1}$. Then, we may form a (quotient) space $\Omega/\mathbb{Z}$ as the equivalent classes in $\Omega$ with the equivalence relation $\omega \sim \psi(\omega)$. To study signals on the joint space $X$ and $\Omega$ with shift $\psi$, we may consider signals on the space $Y=X\times \Omega/\mathbb{Z}$. 

%\begin{Example}
%If $\psi$ is the translation on $\bbR$ as above, then $\bbR/\mathbb{Z}$ is homeomorphic to the circle $S^1$. To study temporal-vertex signals on $X$, we may consider $Y = X\times \bbR/\mathbb{Z} \cong X \times S^1$. A ``smooth signal'' on $Y$ is one that respects both graph topology and periodicity in the time direction. More concretely, if $X$ is a traffic graph, then we may choose $T$ to be 24 hours and $A = \bbR[t]$ or $L^2([0,1])$ to study edge signals. In this case, an edge signal describes how traffic flow changes between sensors during 1 day period.   
%\end{Example}

In summary, we have described two approaches regarding the familiar concept of ``shift'' in signal processing. One is a change-of-coefficient filter that does not alter the simplicial complex. On the other hand, the other requires a change of the simplicial complex by taking the product with $\Omega/\mathbb{Z}$. To deal with a product space arising from the latter approach, we mention that the \emph{Kunneth formula} is a useful tool, details can be found in \cite[p.268]{Hat02}.

%\begin{Theorem}(\cite[p.268]{Hat02}) If $A =\bbR,\mathbb{C}, \mathbb{Z}/2\mathbb{Z}$, we have $H_k(X\times Y,A) \cong \bigoplus_{i+j=k}H_i(X,A)\otimes H_j(Y,A)$. In particular, if $G$ is a graph, then $H_2(X\times S^1) \cong H_1(X)$. 
%\end{Theorem}

\begin{Example}
    \underline{Case study: traffic network.} To illustrate the concepts, we consider a part of the traffic network of Beijing \cite{Lia18}. Road directions are known. Traffic flows over $2$ month, recorded every $15$ mins on $<50\%$ of the roads, are given. 

    We model the network by a directed graph $G_0$. The edge direction follows the road direction in the traffic network. As we only consider a part of the traffic network, we add a node $v_{\infty}$ to represent the remaining (unobserved) network. For each node $v\in G_0$ with only incoming edges, we add an outgoing edge from $v$ to $v_{\infty}$. Similarly, if $v$ only has outgoing edges, we add an edge from $v_{\infty}$ to $v$. The resulting graph is $G$.
    
    Daily periodicity is equivalent to shift invariance w.r.t.\ translation by $T=24$ hrs. We consider $X = G\times S^1$ as described in this subsection, where $S^1$ represents (all $96$ time slots in) $1$ day. To obtain more refined information, we may subdivide (cf.\ \cref{sec:sd}) $X$ by dividing a day into $2$ half days, equivalently, cutting $S^1$ into $2$ edges. As a result, $e\times S^1$ is divided into $4$ triangles, though the topology is unchanged (\figref{fig:fee}). There are $1160$ triangles in total. Let $A$ be the group of polynomials up to degree $5$. Signals are functions $P_{\sigma}\in A$ on the triangles $\sigma$, interpolated from discrete time series. Each such signal represents continuous traffic flows on a (directed) road over a half day. Less than $50\%$ of triangles have their signals known (in the given data). For the task, we want to use $80\%$ of them to predict the rest traffic flow. 

\begin{figure}
    \centering
\includegraphics[scale=0.8]{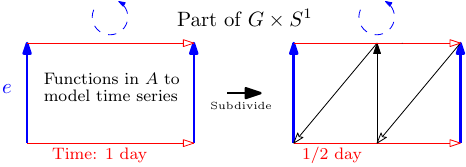}
    \caption{For an edge $e$, subdivision turns $e\times S^1$ into $4$ triangles. A polynomial signal is associated with a triangle accounting for the half-day traffic on $e$.}
    \label{fig:fee}
\end{figure}
As we get access to signals on $<40\%$ triangles, solutions of (\ref{eq:mbb}) are highly non-unique without a regularizer $R(\cdot)$. We assume that for each triangle $\sigma$, there is a rough estimation of the (half day) average traffic flow $a_\sigma$. Then define $R(P_{\sigma}) = \norm{P_{\sigma}-a_{\sigma}}_2$. 

We solve (\ref{eq:mbb}) with this $R(\cdot)$. In \figref{fig:bcg}, we plot the results for signals on $6$ randomly chosen triangles. We have good recovery in many cases. \begin{figure}
        \centering
        \includegraphics[scale=0.45]{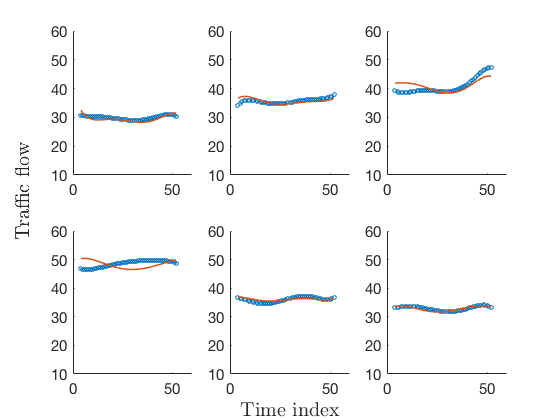}
        \caption{Blue curve: ground truth, red curve: recovered signal.}
        \label{fig:bcg}
    \end{figure}
\end{Example}

\section{Experimental results} \label{sec:exp}

In this section, we use experiments with real datasets to demonstrate the effectiveness of the proposed framework. 

\subsection{Heat flow: bandlimited signals and subdivision} \label{sec:hfb}

In this study, we consider the heat flow dataset of the Intel Berkeley Research lab.\footnote{http://db.csail.mit.edu/labdata/labdata.html} Temperature data are collected from 53 sensors, denoted by $V$, placed in the lab. They form a planar graph $G$ with an edge set $E$ of size $87$. Each edge is given an orientation such that heat flows from high temperature to low temperature. Each (directed) edge is associated with its heat flow estimated from sensor readings at every timestamp. In this way, we obtain the ground truth signals $\{\bx_t, 1\leq t\leq T\}$, where $\bx_t = (x_{e,t})_{e\in E}$ is the (heat flow) edge signals on $G$ at $t$ and $T=864$. For the task, we assume that (randomly chosen) $r \in [0.85,0.95]$ fraction of data is missing and want to estimate all signals from the known $1-r$ fraction. 

\begin{figure}
    \centering
    \includegraphics[height = 3cm,width=4cm]{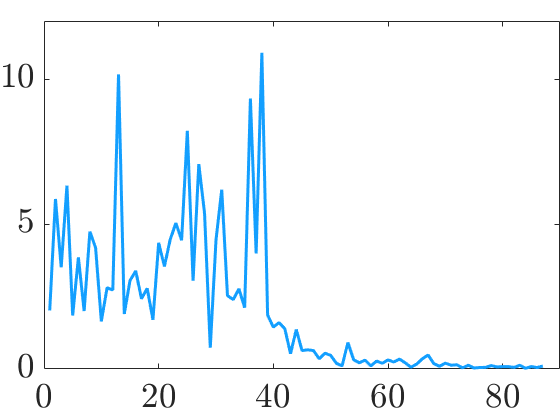}
    \includegraphics[height = 2.85cm, width=4cm]{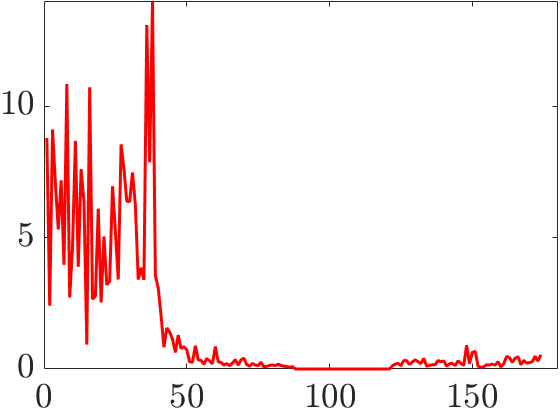}
    \caption{Spectral plots of a typical $\bx_t$ (left) and $\by_t$ (right). Their spectrums are both concentrated for small indices.}
    \label{fig:spo}
\end{figure}

We observe that a typical $\bx_t$ is bandlimited with concentrated spectrum w.r.t.\ $\bL_1$ of $G$ (e.g., \figref{fig:spo}). Motivated by this, we propose the following ``\emph{joint}'' recovery scheme, in the spirit of \cite{Ji19}. Specifically, let $A \subset L^2([0,T/2-1])$ be the group generated by short-time windowed functions \cite{Coh95}: 
\begin{align}
\begin{aligned}\label{eq:wind_funcs}
& \psi_{\omega,t_0}(t) = \exp(-\frac{1}{2\sigma^2}(t-t_0)^2)\cdot\sin(\omega t),\\ & \phi_{\omega,t_0}(t) = \exp(-\frac{1}{2\sigma^2}(t-t_0)^2)\cdot\cos(\omega t),
\end{aligned}
\end{align}
where parameters $t_0 \in T_0=\{0,50,\ldots,400\}$ and $\omega \in \Omega = \{0,1/45,2/45\ldots,1/5\}$. Here, $\sigma$ is a tunable hyperparameter. We use these functions to model temporal patterns of the data. Therefore, a signal at an edge $e$ is a linear combination $f_e = \sum_{t_0\in T_0, \omega \in \Omega}\big(a_{e,\omega,t_0}\psi_{\omega,t_0} + b_{e,\omega,t_0}\phi_{\omega,t_0}\big)$. Write $\boldf = (f_e)_{e\in E}$ for a typical set of edge signals with coefficients in $A$.  

To incorporate the bandlimited spatial information, let $\bF$ be the low-pass filter that projects to space spanned eigenvectors (of $\bL_1$) associated with the smallest $40$ eigenvalues (cf.\ \figref{fig:spo}). Let $\bu_i = (u_{i,e})_{e\in E}$ be the $i$-th eigenvector. Similar to GSP, if $\bg = \bF(\boldf) = (g_e)_{e\in E}$, then $g_e$ can be parameterized by coefficients $\{c_{i,\omega,t_0}, d_{i,\omega,t_0} \mid t_0\in T_0, \omega\in \Omega, i\leq 40\}$ as: 
\begin{align*}
g_e = \sum_{t_0\in T_0, \omega \in \Omega, i\leq 40}u_{i,e}\big(c_{i,\omega,t_0}\psi_{\omega,t_0} + d_{i,\omega,t_0}\phi_{\omega,t_0}\big).
\end{align*}
To solve (\ref{eq:mbb}), we use the fidelity term $\sum_{ (e,t):\text{observed}} \norm{g_e(t)-x_{e,t}}_2^2$ and regularizer $\sum_{t_0,\omega,i}\big(|c_{i,\omega,t_0}|+|d_{i,\omega,t_0}|\big)$. 

The performance is evaluated using the \emph{relative mean squared error (RMSE)} between the estimate $\hat{x}$ and ground truth $x$:
\begin{align*}
    \mathrm{RMSE} := \frac{\sum_{(e,t):\mathrm{test}} (\hat{x}_{e,t} - x_{e,t})^2}{\sum_{(e,t):\mathrm{test}}x_{e,t}^2}
\end{align*}
For comparison, we consider benchmarks in \cite{Coh95} and \cite[(2)]{Rod23}, called ``\emph{seperate}'' and ``\emph{product}'' respectively. Briefly, the separate method performs an interpolation to recover the time-series of each edge based on known signals using the window functions \cref{eq:wind_funcs} as basis. However, spatial signal correlations are not used. On the other hand, the product method construct a new simplical complex $X = G\times P_T$, where $P_T$ is the path graph of size $T$. One can then apply ordinary TSP techinques for recovery on $X$ with signal space $\mathbb{R}$.   

\begin{figure}
    \centering
    \includegraphics[scale = 0.35]{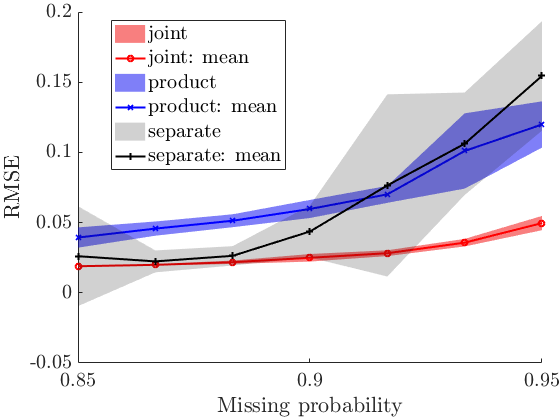}
    \caption{Performance comparison among the joint, separate, and product methods. The curves show average RMSEs over $20$ trials for each $r$, and the shaded regions are derived from sample deviations.}
    \label{fig:pca1}
\end{figure}

The results are shown in \figref{fig:pca1}. We see that our joint method performs significantly better than the benchmarks, particularly for large $r$. Moreover, RMSEs for the joint method have much smaller standard deviations. This suggests that the joint method has fully used both spatial and temporal information. On the other hand, the separate method outperforms the product method for smaller $r$. However, its performance deteriorates if $r$ increases, as it does not use spatial information. 

\begin{figure}
    \centering
    \includegraphics[scale = 0.35]{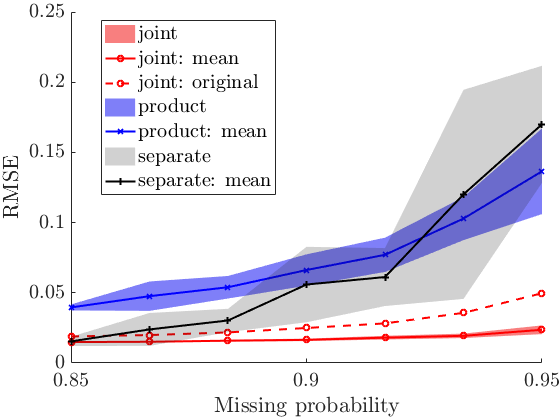}
    \caption{Performance comparison among the joint, separate, and product methods when subdivision is applied. For comparison, we include the ``joint'' curve from \figref{fig:pca1}.}
    \label{fig:pca2}
\end{figure}

Next, we study subdivision. We subdivide each edge of $G$ into two edges and the resulting graph is $G' = (V',E')$. Thus, for each $t$, we have $\by_t = \calE(\bx_t)$ on $E'$ as in \cref{sec:sd}. The spectral plot of a typical $\by_t$ is shown in \figref{fig:spo}. We see that the spectrum is relatively more concentrated, which suggests $\by_t$ is ``smoother'' than $\bx_t$. We perform the same recovery experiments on $G'$. The results are shown in \figref{fig:pca2}. We again see that the joint method outperforms benchmarks with smaller standard deviations. Moreover, comparing with the results for $G$, the RMSE is smaller, particularly for large $r$. This might be due to the fact that the ground truth signals are relatively smoother, as we have observed in \figref{fig:spo}.    

\subsection{Currency exchange: the Hodge decomposition}

In this study, we consider the dataset of currency exchange rates of $9$ countries from 27 Jul.\ 2018 to 26 Jul.\ 2023.\footnote{\url{https://www.ofx.com/en-sg/forex-news/historical-exchange-rates/}} For each pair of currencies, there is a single given exchange rate. For example, between US dollar (USD) and Euro (EUR), only the buying value of 1 USD in EUR is provided. Therefore, if we model the currencies by nodes of a graph $G=(V,E)$, there is a canonical set of edge orientations associated with the dataset. We build a simplicial complex $X$ with edges $E$ and there is a $2$-complex for each triple of nodes.   

Suppose there is no arbitrage. Then for each triple of currencies $v_1, v_2, v_3$, the exchange rates $r_{1,2}, r_{2, 3}$ and $r_{1, 3}$ should satisfy $r_{1, 2}r_{2, 3} = r_{1,3}$. We may turn this into an additive identity on the associated $2$-complex by applying the $\log$ function to the exchange rates. With such preprocessing, we observe edge signals that are $\log$ of exchange rates.  

We assume that on each day, signals on $1/3$ of (uniformly) randomly chosen edges are missing. Therefore, the observed signals are both incomplete and asynchronous. The framework proposed in the paper is suitable to tackle these issues. 

More specifically, for each edge $e = (v_i,v_j) \in E$ and start date $t$, we collect observed signals within $15$ consecutive days starting from $t$. There are usually $<15$ signals due to missing data. However, we can still fit these signals with a polynomial $P_{e,t}$ of degree $2$. Therefore, if we let $A$ be the abelian group of degree $2$ polynomials (over the finite interval $[1, 15]$), $P_{e,t}$ is a generalized signal in $A$. Applying the procedure for all edges, we obtain $\bx_t = \{P_{e,t} \mid e\in E\} \in C_1(X,A)$. 

Endue $A$ with the $L^2$-norm, we perform the optimization version (\ref{eq:mbb}) of the Hodge decomposition. As there is no-nontrivial edge cycle in $X$ (each edge cycle is the boundary of a union of triangles), we have $H_1(X,\mathbb{Z}) = 0$; and hence $H_1(X,A) = 0$ by \cref{thm:iai}. Therefore, we do not need to consider the harmonic component for $\bx_t$. As a result, the solution $\bx_t \approx \bx_t' = \partial_1^*\by_{-1,t}+\partial_2\by_{1,t}, \by_{-1,t} \in C_0(X,A), \by_{1,t} \in C_2(X,A)$. 

As we have discussed, the ``no arbitrage'' condition enforces $\by_{1,t}\approx \bm{0}$. The (generalized) signal  $\by_{-1,t} = \{Q_{v,t} \in A \mid v\in V\}$ is more useful for us to probe changes and trends in the currency exchange market. Intuitively, $Q_{v,t}$ describes the divergence of the signal at node $v$, which can be used to detect change and anomalous behavior. In \figref{fig:ton}, for each $v\in V$, we visualize by plotting the time series $\{Q_{v,t}(1), t \in T\}$ where $T$ is from 27 Jul.\ 2018 to 26 Jul.\ 2023, indexed by 1 - 1819.

For each of the curves, we see that there are noticeable changes, patterns, and trends. We summarize some of the findings and compare them to major real financial and historical events as follows:
\begin{figure*}
\centering
  \includegraphics[scale = 0.33]{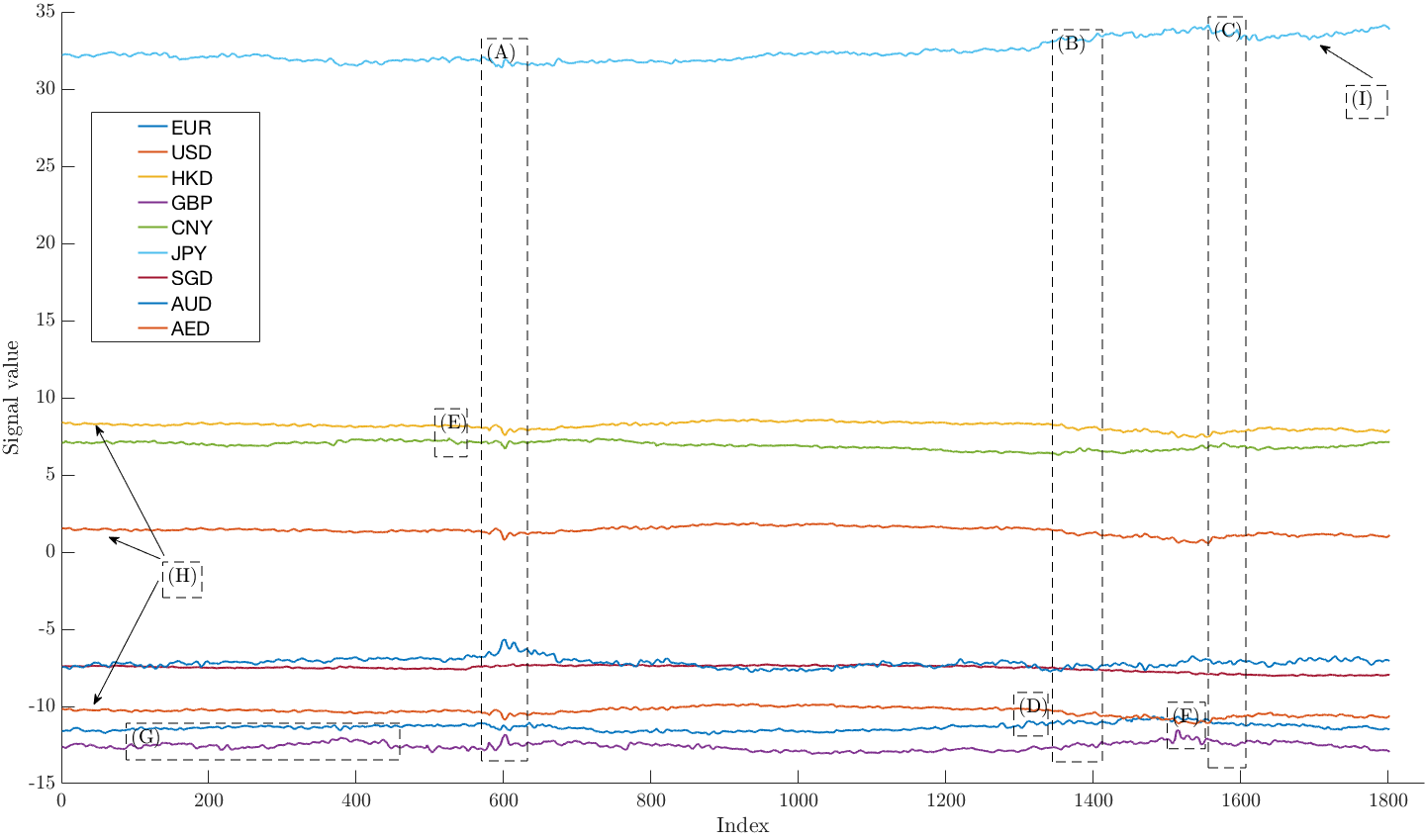}
  \caption{Plots of $Q_{v,t}(1)$ for all nodes corresponding to the nine currencies.} \label{fig:ton}
\end{figure*}

\begin{enumerate}[(A)]
    \item The US Federal Reserve (Fed) is one of the major influencers of the global financial market. Our first few observations are related to Fed policies \cite{Tep23}, which influence most of the large economies. For example, around Index 600  (19 Mar.\ 2020), we see that most curves have a sudden fluctuation. This may correspond to the fact that the Fed reduced the interest rate by 50bps on 3 Mar.\ 2020 and by 100bps on 16 Mar.\ 2020 to combat COVID-19.
    \item Around Index 1340 - 1380 (3 Apr.\ 2022 - 13 May.\ 2022), we see a sudden fluctuation and change in trend for most curves. This may correspond to the fact that the Fed raised the interest rate by 25bps on 17 Mar.\ 2022 and 50bps on 5 May.\ 2022 to combat inflation.
    \item Around Index 1560 - Index 1600 (9 Nov.\ 2022 - 20 Dec.\ 2022), we see a change in the trend for many curves (e.g., AED, HKD, USD from decreasing to increasing, while CNY, JPY, GBP from increasing to decreasing). This may be due to that Fed raised the interest rate by 75bps on 2 Nov.\ 2022, and by 50bps on 14 Dec.\ 2022. It marked the end of the hawkish policies of the Fed. The inflation rate in the US dropped from the highest value of $9.1\%$ to $6.5\%$ by Dec.\ 2022.
    \item For EUR, from around Index 1305 (26 Feb.\ 2022), we start to see frequent fluctuations in the curve. This may correspond to the Russian invasion of Ukraine on 24 Feb.\ 2022.
    \item For the CNY curve, around Index 540 (9 Jan.\ 2020), we see a sudden drop in signal value. This may correspond to the outbreak of COVID-19 at the end of 2019. 
    \item For GBP, around Index 1500 - Index 1550 (10 Sep. 2022 - 30 Oct. 2022), we see many pronounced fluctuations in the curve. The period corresponds to the short premiership of Liz Truss (6 Sep.\ 2022 - 25 Oct.\ 2022). Her economic policies were controversial and resulted in GBP hitting a 37-year low against USD on 23 Sep.\ 2022 \cite{Stu22}.
    \item The GBP curve is unstable around Index 105 - Index 450 (8 Nov.\ 2018 - 21 Oct.\ 2019), as compared with curves of other major economies. This may correspond to the post-Brexit UK-EU negotiation period of the withdrawal agreement (Nov.\ 2018 - 17 Oct.\ 2019). 
    \item AED, HKD, and USD curves are almost identical in shape and trend. One notices that HKD has adopted a linked exchange rate system with USD since Oct.\ 1983, and similarly, AED has been pegged to USD since Nov.\ 1997. 
    \item For the JPY curve, the overall trend is almost increasing for the entire period, even in 2023 (when Fed is less hawkish). This may be due to the Bank of Japan maintaining a negative interest rate of $-0.1\%$ since 2016. 
\end{enumerate}

\begin{figure}
\centering
  \includegraphics[scale = 0.43]{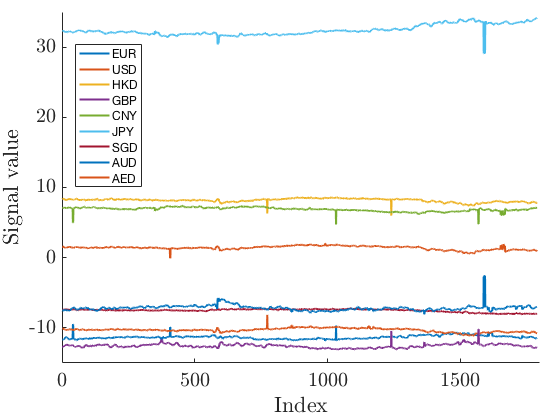}
  \caption{The plots of $\{Q_{v,t}(1), t \in T\}$ if $2/3$ of edge signals are unobserved.} \label{fig:tpi}
\end{figure}

From the observations, we see that the (generalized) Hodge decomposition indeed provides us with useful information, even with incomplete and asynchronous data. In \figref{fig:tpi}, we show the plots of the time series $\{Q_{v,t}(1), t \in T\}$ if $2/3$ edge signals are unobserved. Due to the large amount of missing signals, there are many unusual spikes associated with errors. However, patterns associated with key events, such as the reduction of interest rates by the Fed to combat COVID-19, remain visible. 

\section{Conclusions} \label{sec:conc}
In this paper, we have extended the existing framework of TSP to handle a wider range of signal spaces. We consider observations on simplexes as elements of general abelian groups, so that we may model continuous time series, quantized signals, and asynchronous time series more accurately.  We discuss the decomposition theory of generalized signals. Moreover, we also consider new filter types arising from introducing general signal spaces. Numerical experiments demonstrate the effectiveness of the proposed framework. 

\appendices

\section{Quantized signals} \label{sec:rqs}

For quantized signals, \cref{eq:mbb} is an integer program, which is usually intractable if the simplicial complex is large. A common strategy is to find a continuous approximation. We first introduce the notion of a lattice. 

\begin{Definition}
In the abelian group $\bbR^n$, a subgroup $L$ is a \emph{lattice} if there is a basis $\{\be_1,\ldots, \be_n\}$ of $\bbR^n$ such that $L = \set{\sum_{1\leq i\leq n}a_i\be_i \given a_i\in \mathbb{Z}, 1\leq i\leq n}$, i.e., $L$ consists of the integer linear combinations of the basis. 

The \emph{determinant} $\det(L)$ of $L$ is $|\det(\bE)|$ where $\bE$ is the matrix with columns $\be_i$. The rank of the lattice $L$, denoted by $\rank L$, is $n$. 
\end{Definition}

Recall that a quantized signal on $k$-simplexes of $X$ can be modeled by $C_k(X,\mathbb{Z})$. As $\mathbb{Z}$ is a subset of $\bbR$, we have a natural map $\phi: C_k(X,\mathbb{Z}) \to C_k(X,\bbR)$ (cf.\ the change-of-coefficient filter discussed in more details in \cref{sec:coc}). In particular, it is straightforward to verify that $\phi(\ker \partial_{k,\mathbb{Z}}) \subset \ker \partial_{k,\bbR}$, $\phi(\ima \partial_{k+1, \bbZ}) \subset \ima \partial_{k+1,\bbR}$ and $\phi(\ima \partial^*_{k,\bbZ}) \subset \ima \partial^*_{k,\bbR}$. As $C_k(X,\mathbb{Z})$ is (finitely generated) torsion-free, so are $\ker \partial_{k,\mathbb{Z}}$, $\ima \partial_{k+1,\mathbb{Z}}$ and $\ima \partial^*_{k,\mathbb{Z}}$. We have the following observation.

\begin{Lemma} \label{lem:ppa}
    $\phi(\ker \partial_{k,\mathbb{Z}}), \phi(\ima \partial_{k+1,\mathbb{Z}})$ and $\phi(\ima \partial^*_{k,\mathbb{Z}})$ are lattices in $\ker \partial_{k,\bbR}$, $\ima \partial_{k+1,\bbR}$ and $\ima \partial^*_{k,\bbR}$ respectively.
\end{Lemma}

To solve an optimization in the form of (\ref{eq:mbb})  for quantized signals, we may perform the following approximations:
\begin{itemize}
    \item Solve the same problem in real numbers to obtain $\widehat{\bx}$.
    \item Find the closest point of $\widehat{\bx}$ in the embedded lattice (\cref{lem:ppa}) to obtain $\widehat{\bx}_p$ or round off the components of $\widehat{\bx}$ to the nearest integers to obtain $\widehat{\bx}_r$. 
\end{itemize}

Finding $\widehat{\bx}_p$ is the closest vector problem (CVP), which is usually computationally costly. On the other hand, $\widehat{\bx}_r$ is not always a feasible solution. Therefore, the exact approach to adopt should be determined by specific applications. We can also estimate the quality of the approximated solution.  

\begin{Lemma} \label{lem:sto}
Suppose the optimization takes the form (cf.\ (\ref{eq:mbb}))
\begin{align*}
    \min_{(\bx_0 ,\bx_1,\bx_{-1})} & f(\bx') \\
    \ST & \bx_0 + \bx_1 + \bx_{-1} = \bx'\\
    & \bx_0 \in \ker\partial_{k,\mathbb{Z}},\\
    & \bx_1 = \partial_{k+1,\mathbb{Z}}(\by_1), \by_1\in C_{k+1}(X,\mathbb{Z}),\\
    & \bx_{-1} = \partial_{k,\mathbb{Z}}^*(\by_{-1}), \by_{-1}\in C_{k-1}(X,\mathbb{Z}).
\end{align*}
Let $\widehat{\bx}_p, \widehat{\bx}_r$ be as above, and $\widehat{\bx}_e$ be the exact solution to the integer problem. For convenience, denote $\ker\partial_{k,\mathbb{Z}}, \ima \partial_{k+1,\mathbb{Z}}$ and $\ima \partial_{k,\mathbb{Z}}^*$ by $S_1, S_2$ and $S_3$, respectively. If $f$ is $\alpha$-Lipschitz w.r.t.\ the $L^\infty$ norm, then we have
\begin{align*}
    \max (f(\widehat{\bx}_r)-\alpha/2, f(\widehat{\bx}_p)-\alpha\beta_p)\leq f(\widehat{\bx}_e) \leq f(\widehat{\bx}_p),  
\end{align*}
where $\beta_p = \max\set{|\det(S_i)|^{1/\rank S_i}\given i=1,2,3}$. 
\end{Lemma}

\begin{Example}
    \underline{Case study: coauthorship complex.} The work \cite{Ebl20} considers a simplicial complex $X$ based on \cite{Pat17}, whose nodes correspond to authors. A $k$-simplex $\sigma^k$ is formed for $k+1$ collaborating authors. The signal assigned to each $k$-simplex $\sigma^k$ is the number of citations attributed to collaborations among the $k+1$ authors associated with $\sigma^k$, which belongs to $\mathbb{Z}$. The imputation task (cf.\ \cite{Ebl20}) requires one to estimate missing signals if only partial observations are made. 

    We consider edge signal imputation assuming $r\%$ of signals are missing. Denote the observed signals by $\bz = \bP(\bx)$, where $\bx$ is the ground truth and $\bP$ is the projection to the signal space on edges with observations. For the task, we estimate $\widehat{\bx}$ by solving (\ref{eq:mbb}) with the following objective function: The fidelity term $\norm{\bx'-\bx}_{k,2}$ is replaced by $\norm{\bP(\bx')-\bz}_{k,2}$. For the regularizer $R(\cdot)$, we construct a (multi-)graph $G_X$, whose nodes are edges of $X$. Two nodes in $G_X$ are connected by an edge if they share a common node in $X$ or they are on the boundary of the same $2$-complex. Hence, there can be two edges between a pair of nodes in $G_X$. Let $\bL_{G_X}$ be its Laplacian. Then $R(\bx')$ is defined as $\langle \bx', \bL_{G_X}\bx'\rangle$. It regularizes signal smoothness for ``adjacent'' edges.  

    Based on the discussion after \cref{lem:ppa}, we obtain an integer signal $\widehat{\bx}_r$ by rounding off $\widehat{\bx}$. The imputation error distributions for $40\%$ and $50\%$ of missing data are shown in \figref{fig:edf}. The accuracies are $\approx 76.6\%$ and $ 75.3\%$ respectively, which are higher than those of SNN in \cite{Ebl20} ($\approx 66\%$ and $50\%$, respectively). 
    \begin{figure}
        \centering
        \includegraphics[scale=0.25]{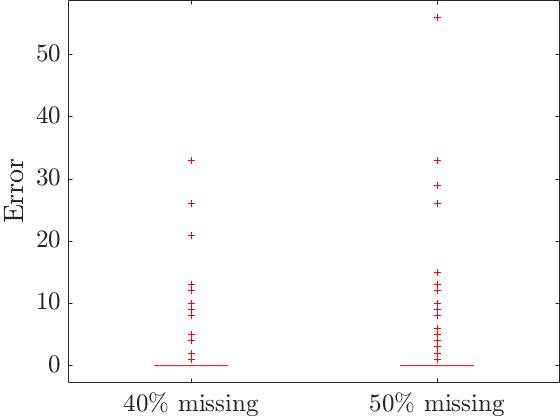}
        \caption{Boxplots of error distributions for imputation.}
        \label{fig:edf}
    \end{figure}
\end{Example}

\section{Approximated band-pass filters} \label{sec:abf}
In this subsection, we demonstrate how the fundamental learning framework can approximate algebraic procedures such as band-pass filters w.r.t.\ the Hodge Laplacian. 

Suppose for an ordinary graph signal $\bx$ on a connected graph, the Hodge decomposition yields $\bx=\bx_0+\bx_1$ with $\bx_0$ the harmonic component. On the other hand, GSP offers an eigendecomposition $\bx$ into the eigenspaces of the graph Laplacian $\bL_0$ (called \emph{the Laplacian decomposition}), and $\bx_0$ corresponds to the component of the constant eigenvector. However, this type of spectral analysis cannot be applied for a general coefficient $A$, as eigendecomposition is not always well-defined.   

To motivate how (\ref{eq:mbb}) might be used, we examine the Laplacian decomposition in GSP from a different perspective. Recall that the Courant-Fischer theorem %\cite{Tao12} 
states that if $V_{l-1} = \{\be_1,\ldots,\be_{l-1}\}$ are the eigenvectors of a symmetric matrix $\bM$ with the smallest eigenvalues. Then the $l$-th smallest eigenvalue and its associated unit eigenvector $\be_l$ can be determined by
\begin{align*}
    \lambda_l = \langle \be_l, \bM \be_l\rangle =  \min_{0\neq \bx\perp \text{ span}(V_{l-1})}\frac{\langle \bx, \bM \bx\rangle}{\langle \bx, \bx \rangle}.
\end{align*}

In GSP, the Laplacian $\bL_0 = \bB_1\bB_1^*$ is used. If the graph is connected, for $l>1$, $\be_l$ does not have any harmonic component and $\be_l = \bB_1\by_l$ for some edge signal $\by_l$. Therefore, 
\begin{align} \label{eq:llb}
    \begin{split}
   \lambda_l = \langle \be_l, \bL_0 \be_l\rangle = \frac{\langle \bB_1^*\bB_1\by_l, \bB_1^*\bB_1 \by_l\rangle}{\langle \bB_1\by_l, \bB_1\by_l\rangle}.       
    \end{split}
\end{align}
The formula suggests that algebraic quantities such as the eigenvalues of the Laplacian can also be retrieved by optimization. Though for a general $A$, we do not necessarily have a theory of eigendecomposition of the Laplacians, the right-hand side (RHS) of (\ref{eq:llb}) still makes sense. Thus for $\bx$ without the harmonic component (which can be removed by the generalized Hodge decomposition), we are inspired to set up a special form of (\ref{eq:mbb}) as follows: 
\begin{align} 
\begin{aligned}\label{eq:mbn}
    \min_{\bx'} &\norm{\bx' -\bx}_{k,p}^p + \frac{1}{\eta} \big(\norm{\partial_{k+1}^*\bx_1}_{k+1,p}^p+\norm{\partial_k\bx_{-1}}_{k-1,p}^p\big)\\
    \ST & \bx' = \bx_1 + \bx_{-1}, \\
    & \bx_1 = \partial_{k+1}(\by_1), \by_1\in C_{k+1}(X,A),\\
    & \bx_{-1} = \partial_k^*(\by_{-1}), \by_{-1}\in C_{k-1}(X,A).
\end{aligned}
\end{align}

The following observation justifies the formulation. 

\begin{Proposition} \label{thm:sam}
    Suppose $A = \mathbb{R}$ and $p=2$. Let $\lambda_1\leq \ldots\leq \lambda_{n_k}$ be an ordered sequence of eigenvalues of $\bL_k$ and their associate eigenvectors are $\be_1,\ldots, \be_{n_k}$. For $\bx \in C_k(X,A)$, assume its eigendecomposition is $\bx = \sum_{1\leq i\leq n_k} a_i\be_i, a_i\in A$. Then for $\bx' = \sum_{1\leq i\leq n_k} b_i\be_i$, if $\bx'$ is solution to (\ref{eq:mbn}), then the following inequalities hold:
    \begin{align}
    |b_i| \leq \frac{2|a_i|}{1+\frac{\lambda_i}{\eta}} \text{ and }
    |b_i-\frac{a_i}{1+\frac{\lambda_i}{\eta}}| \leq \frac{|a_i|}{1+\frac{\eta}{\lambda_i}},
    \end{align}
for any $1\leq i\leq n_k$.
\end{Proposition}

The result essentially says that for a solution $\bx' = \sum_{1\leq i\leq n_k} b_i\be_i$ and some $1\leq i\leq n_k$, if $\lambda_i \gg \eta$, then $b_i$ must be close to $0$. On the other hand, if $\lambda_i \ll \eta$, then $b_i$ must be close to $a_i$. This indicates that solving (\ref{eq:mbn}) is approximately a band-pass filter. 

\section{Proofs}

If we fix $A = \bbR$, then $C_k(X, A)$ is a finite-dimensional vector space. We first provide a self-contained study of relations among various subgroups of $C_k(X, A)$. 

For $C_k(X,A)$, there are a few subspaces we are interested in: $\ima\partial_{k+1}, \ima\partial_k^*, \ker\partial_k, \ker\partial_{k+1}^*$ and $\ker\bL_k$. We have the following observations:

\begin{enumerate}[(a)]
    \item \emph{$\ima\partial_{k+1} \subset \ker\partial_k$}: This is a due to $\partial_k\circ \partial_{k+1}=0$.
    \item \emph{$\ima\partial_k^* \subset \ker \partial_{k+1}^*$}: This due to $\partial_{k+1}^*\circ \partial_k^*=0$.
    \item \emph{$\ima\partial_{k+1} \perp \ima\partial_k^*$}: For any $\bx=\partial_{k+1}\by$ and $\bx'=\partial_k^*\by'$, we have $\langle \bx,\bx' \rangle = \langle \partial_{k+1}\by, \partial_k^*\by'\rangle = \langle \partial_k\circ \partial_{k+1}\by,\by'\rangle =0$. Therefore, $\ima\partial_{k+1}$ and $\ima\partial_k^*$ are orthogonal. 
    \item \label{it:kpi} $\ker\partial_k = (\ima\partial_k^*)^{\perp}$: As $\rank \partial_k^* + \dim \ker\partial_k = \dim C_k(X,A)$, it suffices to show that $\ker\partial_k \perp \ima\partial_k^*$. For $\bx \in \ker\partial_k$ and $\partial_k^* \by \in \ima\partial_k^*$, we have $\langle \bx, \partial_k^* \by\rangle = \langle \partial_k\bx, \by\rangle=0$.
    \item $\ker \bL_k \subset \ker\partial_k \cap \ker \partial_{k+1}^*$: For any $\bx \in \ker \bL_k$, we have
    \begin{align*}
       0 = \langle \bL_k \bx, \bL_k \bx\rangle 
       & = \langle \partial_k^*\circ \partial_k \bx, \partial_k^*\circ \partial_k \bx\rangle \\ &+ \langle \partial_{k+1}\circ \partial_{k+1}^* \bx, \partial_{k+1}\circ \partial_{k+1}^* \bx\rangle.
    \end{align*}
    Therefore, $\partial_k^*\circ \partial_k \bx = 0$ and $\partial_{k+1}\circ \partial_{k+1}^* \bx=0$. This implies that $0 = \langle \partial_k^*\circ \partial_k \bx, \bx \rangle = \langle \partial_k\bx,\partial_k\bx\rangle$ and hence $\partial_k\bx = 0$. Similarly, $\partial_{k+1}^*\bx =0$.
    \item $\ker \bL_k = (\ima\partial_{k+1}\oplus \ima\partial_k^*)^{\perp}$: For any $\bx \in \ker\bL_k$ and $\by \in C_{k+1}(X,A)$, we have 
    \begin{align*}
        \langle \bx, \partial_{k+1}\by \rangle = \langle \partial_{k+1}^*\bx,\by \rangle = 0.
    \end{align*}
    Hence, $\ker\bL_k \perp \ima\partial_{k+1}$ and similarly,
$\ker\bL_k \perp \ima\partial_k^*$. This means $\ker \bL_k \subset (\ima\partial_{k+1}\oplus \ima\partial_k^*)^{\perp}$. 

Conversely, if $\bx \in (\ima\partial_{k+1}\oplus \ima\partial_k^*)^{\perp}$, we compute
\begin{align*}
    & \langle \bL_k\bx,\bL_k\bx \rangle \\
 = &\langle \partial_k^*\circ \partial_k \bx, \partial_k^*\circ \partial_k \bx\rangle + \langle \partial_{k+1}\circ \partial_{k+1}^* \bx, \partial_{k+1}\circ \partial_{k+1}^* \bx\rangle \\
= & \langle \bx, \partial_k^*\circ \partial_k \circ \partial_k^*\circ \partial_k \bx\rangle \\
& + \langle \bx, \partial_{k+1}\circ \partial_{k+1}^* \circ \partial_{k+1}\circ \partial_{k+1}^* \bx\rangle = 0. 
\end{align*}
Therefore, $(\ima\partial_{k+1}\oplus \ima\partial_k^*)^{\perp} \subset \ker \bL_k$ and The claim follows. This is essentially the Hodge decomposition.
\item $\ker\partial_k = \ima\partial_{k+1}\oplus \ker \bL_k$: it follows from \ref{it:kpi} and the Hodge decomposition.
\end{enumerate}

We are ready to prove the results of the paper. 

\begin{IEEEproof}[Proof of \cref{lem:fam}]
This result is partially known and we give a self-contained complete proof. If $A=\bbR$, it suffices to show that $\dim H_k(X, A) = \dim \ker \bL_k$ as they are finite dimensional vector spaces. By the above observations, we have 
\begin{align*}
    \rank \partial_{k+1} + \rank \partial_k + \dim \ker \bL_k = \dim C_k(X,A), 
\end{align*}
using $\rank \partial_k = \rank \partial_k^*$.

Moreover, $\dim H_k(X,A) + \rank \partial_{k+1} = \dim \ker\partial_k$ and $\dim \ker\partial_k + \rank \partial_k = \dim C_k(X,A)$. Summing up these identities and cancelling $\dim \ker\partial_k$, we have 
\begin{align*}
        \rank \partial_{k+1} + \rank \partial_k + \dim H_k(X,A) = \dim C_k(X,A),
\end{align*}
and the result follows.

If $A = \mathbb{Z}$, by the same argument as (e) and (f) above, we have $\ker \bL_k \subset \ker \partial_k$ and $\ker \bL_k \perp \ima \partial_{k+1}$. Hence, $\ker \bL_k \cap \ima \partial_{k+1} = 0$. This implies that $\ker \bL_k \subset \ker \partial_k/\ima \partial_{k+1} = H_k(X,A)$. 

In $A$ is torsion free, we have $\ker \bL_{k,A} \cong \bL_{k,\mathbb{Z}}\otimes A$ \cite{Lan02} p.612. Similarly, if $A$ is an $\bbR$-vector space, we have $\ker \bL_{k,A} \cong \bL_{k,\bbR}\otimes A$. Therefore, the general cases follow from \cref{thm:iai} and the cases for $A = \mathbb{Z},\bbR$.  
\end{IEEEproof}

\begin{IEEEproof}[Proof of \cref{thm:fam}]
    For $\bx \in C_k(X,A)$, assume that its Hodge decomposition is $\bx = \tilde{\bx}_0 + \tilde{\bx}_1 + \tilde{\bx}_{-1}, \tilde{\bx}_0 \in \ker \bL_k \subset \ker\partial_k, \tilde{\bx}_1\in \ima\partial_{k+1}, \tilde{\bx}_{-1} \in \ima\partial_k^*$. 

    We first consider the case $\bx'=\bx$ and (\ref{eq:mbb}) requires us to minimize $\norm{\bx_0}_2$. The space $C_k(X,A)$ has orthogonal decomposition $\ima\partial_k^* \oplus \ker\partial_k$. Therefore, to minimize $\norm{\bx_0}_2$, we first need to project $\bx$ to $\ker\partial_k$. Moreover, $\ker\partial_k = \ima\partial_{k+1}\oplus \ker \bL_k$. Note that we are allowed to remove a summand in $\ima\partial_{k+1}$ from $\bx$ to obtain $\bx_0$ with the smallest norm. Therefore, to minimize $\norm{\bx_0}_2$, the summand we need to remove from $\bx$ is its projection to $\ima\partial_{k+1}$. Therefore, the Hodge decomposition is the unique solution to (\ref{eq:mbb}) provided $\bx'=\bx$.

    To conclude, it suffices to show that if $\bx'\neq \bx$. Then 
    \begin{align*}
    \norm{\tilde{\bx}_0}_2< \norm{\bx_0}_2 + \zeta \norm{\bx'-\bx}_{k,2}.   
    \end{align*}
    By the orthogonality of the different components of the Hodge decomposition 
    \begin{align*}
    \norm{\bx'-\bx}_{k,2}^2 & =  \norm{\bx_1-\tilde{\bx}_1}_{k,2}^2 \\ & + \norm{\bx_{-1}-\tilde{\bx}_{-1}}_{k,2}^2 + \norm{\bx_0-\tilde{\bx}_0}_2^2.
    \end{align*}
    Therefore, $\norm{\bx'-\bx}_{k,2} \geq \norm{\bx_0-\tilde{\bx}_0}_2$. By the triangle inequality, we estimate that 
    \begin{align*}
    \norm{\tilde{\bx}_0}_2 &\leq \norm{\bx_0}_2 + \norm{\bx_0-\tilde{\bx}_0}_2 \leq \norm{\bx_0}_2 + \norm{\bx'-\bx}_2 \\
    & < \norm{\bx_0}_2 + \zeta \norm{\bx'-\bx}_2,  
    \end{align*}
    and the result follows.
\end{IEEEproof}

\begin{IEEEproof}[Proof of \cref{lem:ibb}]
The result follows immediately from the observation that $\mathcal{E}(\partial_{k+1}\by) = \partial_{k+1}\mathcal{E}(\by)$, where $\partial_{k+1}$ are for $X$ and $Y$ respectively. It suffices to show the identity for $\by$ a basis element, i.e., supported on a single simplex. This case follows from our choice of the orientation that the internal simplexes of $L_1,\ldots, L_{m_k}$ have opposite orientations. Therefore, their contribution in $\partial_{k+1}(\by)$ are cancelled. The boundary of $\mathcal{E}(\by)$ is the subdivision of the boundary of $\by$ as claimed. Similarly, we also have $\mathcal{E}(\partial_k^*\bz) = \partial_k^*(\mathcal{E}(\bz))$.

For the contractive property, $\norm{\mathcal{E}(\bx)_0}_{Y,k,p} \leq \norm{\mathcal{E}(\bx_0)}_{Y,k,p}$ follows from the fact that $\mathcal{E}(\bx)_0$ is (part of) a solution of (\ref{eq:mbb}) for $\mathcal{E}(\bx)$. The identity $\norm{\mathcal{E}(\bx_0)}_{Y,k,p} = \norm{\bx_0}_{X,k,p}$ follows from the definition of $\norm{\cdot}_{Y,k,p}$ and $\norm{\cdot}_{X,k,p}$ and our choice of the subdivision of $\bw$ that perserves the $p$-norm.
\end{IEEEproof}

\begin{IEEEproof}[Proof of \cref{lem:pia}]
    Notice that $\phi_{*,k-1}\circ \partial_{k,A} = \partial_{k,A'}\circ\phi_{*,k}$. For $\bx \in \ker\partial_{k,A}$, we have $\partial_{k,A'}\circ\phi_{*,k}(\bx) = \phi_{*,k-1}\circ \partial_{k,A}(\bx) =0$ and $\phi_{*,k}(\bx) \in \ker\partial_{k,A'}$. Hence, $\phi_{*,k}(\ker \partial_{k,A})\subset \ker \partial_{k,A'}$. The other two set inclusions can be proved similarly.
\end{IEEEproof}

\begin{IEEEproof}[Proof of \cref{lem:ppa}]
This is a standard result in homological algebra. Precisely, it is because $\bbR$ is torsion-free and hence flat over $\mathbb{Z}$ \cite[p.\ 612]{Lan02}. 
\end{IEEEproof}

\begin{IEEEproof}[Proof of \cref{lem:sto}]
As $\widehat{\bx}_p$ is a feasible integer solution to the optimization problem. We have $f(\widehat{\bx}_e) \leq f(\widehat{\bx}_p)$.     

As $\widehat{x}$ is a $\bbR$ solution of the optimization problem. We have $f(\widehat{\bx}) \leq f(\widehat{\bx}_e)$. Moreover, $\norm{\widehat{\bx}-\widehat{\bx}_r}_{\infty} \leq 1/2$, the Lipschitz condition of $f$ implies that $\widehat{\bx}_r-\alpha/2 \leq f(\widehat{\bx}_e)$. 

On the other hand, \cref{lem:ppa} implies that $S_1, S_2$ and $S_3$ are lattices in the vector spaces $\ker\partial_{k,\bbR}, \ima \partial_{k+1,\bbR}$ and $\ima \partial_{k,\bbR}^*$, respectively. By Minkowski's theorem \cite{Cas12}, we have $\norm{\widehat{\bx}-\widehat{\bx}_p}_{\infty} \leq \beta_p$. Hence, by the Lipschitz condition of $f$, $\widehat{\bx}_p-\alpha\beta_p \leq f(\widehat{\bx}_e)$.
\end{IEEEproof}

\begin{IEEEproof}[Proof of \cref{thm:sam}]
Denote 
\begin{align*}\norm{\bx' -\bx}_{k,2}^2 + \frac{1}{\eta} \big(\norm{\partial_{k+1}^*\bx_1}_{k+1,2}^2+\norm{\partial_k\bx_{-1}}_{k-1,2}^2\big)\end{align*} 
by $\psi_{\bx}(\bx')$. Suppose for some $1\leq i\leq n_k$, we have 
\begin{align*}
    |b_i| > \frac{2|a_i|}{1+\frac{\lambda_i}{\eta}}.
\end{align*}

Let $\bx'' = \bx' - b_i\be_i \neq \bx'$, i.e., we replace $b_i$ by $0$. As we only change the $i$-th eigen-component, $\norm{\bx''-\bx}_{k,2}^2-\norm{\bx'-\bx}_{k,2}^2$ is $a_i^2-(a_i-b_i)^2=2a_ib_i-b_i^2$. On the other hand, in general for any $\bx = \bx_1+\bx_{-1}$, we notice that 
    \begin{align*}
& \norm{\partial_{k+1}^*\bx_1}_{k+1,2}^2+\norm{\partial_k\bx_{-1}}_{k-1,2}^2 \\ =& \langle \partial_{k+1}^*\bx_1, \partial_{k+1}^*\bx_1 \rangle + \langle \partial_k\bx_{-1},\partial_k\bx_{-1}\rangle \\ = & \langle \bx_1, \partial_{k+1}\circ \partial_{k+1}^*\bx_1\rangle + \langle \bx_{-1}, \partial_k^*\circ \partial_k\bx_{-1}\rangle \\ =& \langle \bx_1+\bx_{-1}, \partial_{k+1}\circ \partial_{k+1}^*\bx_1\rangle + \langle \bx_1+\bx_{-1}, \partial_k^*\circ \partial_k\bx_{-1}\rangle\\=& \langle \bx, \partial_{k+1}\circ \partial_{k+1}^*\bx_1+\partial_k^*\circ \partial_k\bx_{-1} \rangle \\ = & \langle \bx, \partial_{k+1}\circ \partial_{k+1}^*(\bx_1+\bx_{-1})+\partial_k^*\circ \partial_k(\bx_1+\bx_{-1}) \rangle\\= &\langle \bx, \bL_k \bx\rangle.  
    \end{align*}
As $\bx'$ does not have any harmonic component, the same holds for $\bx''$. If $\bx'' = \bx_1''+\bx_{-1}''$, then 
\begin{align*}
& \norm{\partial_{k+1}^*\bx_1'}_{k+1,2}^2+\norm{\partial_k\bx_{-1}'}_{k-1,2}^2 \\ & -(\norm{\partial_{k+1}^*\bx_1''}_{k+1,2}^2+\norm{\partial_k\bx_{-1}''}_{k-1,2}^2) \\ = & \langle \bx', \bL_k \bx'\rangle - \langle \bx'', \bL_k \bx''\rangle \\
= & b_i^2\langle \be_i, \lambda_i\be_i \rangle = \lambda_ib_i^2. 
\end{align*}
Therefore 
\begin{align*}
&\psi_{\bx}(\bx'') - \psi_{\bx}(\bx')
= 2a_ib_i-b_i^2 - \lambda_ib_i^2/\eta \\= & b_i\big(2a_i-(1+\frac{\lambda_i}{\eta})b_i\big).
\end{align*}
This value is negative by the assumption that
\begin{align*}
    |b_i| > \frac{2|a_i|}{1+\frac{\lambda_i}{\eta}}.
\end{align*}
Hence, $\bx'$ cannot be a solution of (\ref{eq:mbn}), which is a contradiction. 

For the second part, if 
\begin{align} \label{eq:ba1}
|b_i-\frac{a_i}{1+\frac{\lambda_i}{\eta}}| > \frac{|a_i|}{1+\frac{\eta}{\lambda_i}}, 
\end{align}
we form $\bx''$ by replacing $b_i$ with $a_i$. By the same calculation as above, we have $\norm{\bx''-\bx}_{k,2}^2-\norm{\bx'-\bx}_{k,2}^2 = -(a_i-b_i)^2$, and  
\begin{align*}
& \norm{\partial_{k+1}^*\bx_1'}_{k+1,2}^2+\norm{\partial_k\bx_{-1}'}_{k-1,2}^2 \\ & -(\norm{\partial_{k+1}^*\bx_1''}_{k+1,2}^2+\norm{\partial_k\bx_{-1}''}_{k-1,2}^2) \\
= & \lambda_i(b_i^2-a_i^2).  
\end{align*}
Therefore, 
\begin{align*}
&\psi_{\bx}(\bx'') - \psi_{\bx}(\bx')
=  \frac{\lambda_i}{\eta}(a_i^2-b_i^2)-(a_i-b_i)^2 \\
= & - (b_i^2 - \frac{2a_i}{1+\frac{\lambda_i}{\eta}} b_i + \frac{1-\frac{\lambda_i}{\eta}}{1+\frac{\lambda_i}{\eta}}a_i^2)(1+\frac{\lambda_i}{\eta}).
\end{align*}
For this quadratic expression, if (\ref{eq:ba1}) holds, it is always negative. Therefore, $\bx'$ cannot be a solution of (\ref{eq:mbn}), which is a contradiction. The result is proved. 
\end{IEEEproof}

\bibliographystyle{IEEEtran}
\bibliography{IEEEabrv,StringDefinitions,allref}

\end{document}